\documentclass[a4, 11pt]{article}

\usepackage{fullpage}

\usepackage[utf8]{inputenc}
\usepackage[T1]{fontenc}

\usepackage{amssymb}
\usepackage{latexsym}
\usepackage{array}

\newcommand{\eps}{\epsilon}

\usepackage{graphicx}
\usepackage{algorithm}
\usepackage{algpseudocode}

\usepackage{tikz}
\usetikzlibrary{decorations.pathmorphing}
\usetikzlibrary{positioning,shapes,calc}
\usepackage{latawce}
\usepackage{subcaption}
\usepackage{authblk}

\title{A $4/5$ - Approximation Algorithm  for the Maximum  Traveling Salesman Problem\thanks{Partly supported by Polish National Science Center grant UMO-2013/11/B/ST6/01748}}

\author{Szymon Dudycz\thanks{szymon.dudycz@gmail.com}, \enskip
		Jan Marcinkowski\thanks{jasiekmarc@stud.cs.uni.wroc.pl}, \enskip 
		Katarzyna Paluch\thanks{abraka@cs.uni.wroc.pl},\enskip and \enskip
		Bartosz Rybicki\thanks{rybicki.bartek@gmail.com}}

\affil{Institute of Computer Science,  University of Wroc{\l}aw}

\date{}

\newcommand{\dowod}{\noindent{\bf Proof.~}}
\newcommand{\koniec}{\hfill $\Box$\\[.1ex]}

\renewcommand{\>}{\rangle}
\newtheorem{fact}{Fact}

\newcommand{\Kt}{{\cal K}_3}
\newcommand{\Kd}{{\cal K}_2}

\newtheorem{lemma}{Lemma}
\newtheorem{theorem}{Theorem}
\newtheorem{corollary}{Corollary}
\newtheorem{definition}{Definition}

\newtheorem{invariant}{Invariant}

\begin{document}

\maketitle
\thispagestyle{empty}
\begin{abstract}
In the maximum  traveling salesman problem (Max TSP)  we are given a complete undirected graph with nonnegative weights on the edges and we wish to compute a traveling salesman tour of maximum weight. We present a fast combinatorial  $\frac 45$ - approximation algorithm for Max TSP.
The previous best approximation for this problem was $\frac 79$. The new algorithm is based on a novel technique of eliminating difficult subgraphs via {\em half-edges}, a new method of edge coloring and a technique of exchanging edges. A {\it half-edge} of edge $e=(u,v)$ is informally speaking ``a half of $e$ containing either $u$ or $v$''.

\end{abstract}

\newpage

\section{Introduction}
The maximum  traveling salesman problem (Max TSP) is a classical variant of the
famous traveling salesman problem. In the problem we are given a complete undirected graph
$G=(V,E)$ with nonnegative weights on the edges and we wish to compute a
traveling salesman tour of maximum weight. Max TSP, also informally known as the ``taxicab
ripoff problem'' is both of theoretical and practical interest.

Previous approximations of  Max
TSP have found applications in combinatorics and computational biology: the problem is useful in understanding RNA
interactions~\cite{RNA} and providing algorithms for compressing the results of
DNA sequencing~\cite{DNASEQ}. It has also been  applied to a problem of finding
a maximum weight triangle cover of the graph~\cite{HRTri} and to a combinatorial
problem called \emph{bandpass-2}~\cite{CW}, where we are supposed to find the
best permutation of rows in a boolean-valued matrix, so that the weighted sum of
structures called \emph{bandpasses} is maximised.

{\bf Previous results.}  The first approximation algorithms for Max TSP were devised by Fisher, Nemhauser and Wolsey \cite{Fish}. They showed several algorithms having approximation ratio $\frac {1}{2}$
and  one  having a guarantee of $\frac{2}{3}$. In \cite{Kos} Kosaraju, Park and Stein  presented an improved algorithm having a  ratio $\frac{19}{27}$  (\cite{BH}). This was in turn improved by Hassin and Rubinstein, who gave a $\frac 57$- approximation (\cite{HR1}).
In the meantime  Serdyukov \cite{Ser} presented (in Russian) a simple and elegant $\frac{3}{4}$-approximation algorithm. The algorithm
is deterministic and runs in $O(n^3)$, where $n$ denotes the number of vertices in the graph.
Afterwards, Hassin and Rubinstein (\cite{HR}) gave a randomized algorithm having  expected approximation ratio at least $\frac{25(1-\eps)}{33-32\eps}$  and running in $O(n^2(n+2^{1/\eps}))$, where $\eps$ is an arbitrarily small constant.
The first deterministic approximation algorithm with the ratio
better than $\frac{3}{4}$ was given in \cite{Chen} by Chen, Okamoto and  Wang. It is a $\frac{61}{81}$-approximation and a nontrivial derandomization of the algorithm from \cite{HR} and runs in $O(n^3)$. The currently best known approximation has been given by Paluch, Mucha and Madry \cite{Paluch}  and achieves the ratio of $\frac 79$. Its running time  is also $O(n^3)$.

{\bf Related Work.} It is known that Max-TSP  is max-SNP-hard \cite{bgww}, so there exists a constant $\delta < 1$, which is an upper bound on the approximation ratio of any algorithm for this problem. The geometric version of the problem, where all vertices are in $R^d$ and the  weight of each edge is defined as the Euclidean distance of its endpoints, was considered in \cite{g_maxtsp}. The algorithm presented in this paper solves the problem exactly in polynomial time, assuming that the number $d$ of dimensions is constant. Moreover, it is quite fast for real-life instances, in which $d$ is small. 

Regarding the path version of Max TSP - Max-TSPP (the Maximum Traveling Salesman Path Problem), the approximation algorithms with ratios correspondingly $\frac 12$ and $\frac 23$ have been given in \cite{Monnot}. The first one is for the case when both endpoints of the path are specified and the other for the case when only one endpoint is given.

Another related problem is called the maximum scatter TSP (see \cite{arkin}). In it the goal is to find a TSP tour (or a path) which maximizes the weight of the minimum weight (lightest) edge selected in the solution. The problem is motivated by medical imaging and some manufacturing applications.  In general there is no constant approximation for this problem, but if  the weights of the edges obey the triangle inequality, then it is possible to give a $\frac{1}{2}$-approximation  algorithm. The paper studies also the more general version of the maximum scatter TSP -- the max-min-$m$-neighbour TSP. The improved approximation results  for the max-min-$2$-neighbour problem have been given in \cite{Chiang}.

In the maximum latency TSP problem  we are given a complete undirected graph with vertices $v_0, v_1, \ldots, v_n$. Our goal is to find a Hamiltonian path starting at a fixed vertex $v_0$, which maximizes the total latency of the vertices. If in a given path $P$ the weight of the $i$-th edge is $w_i$, then the latency of the $j$-th vertex is $L_j = \sum_{i = 1}^j w_i$ and the total latency is defined as $L(P) = \sum_{j = 1}^n L_j$. A ratio $\frac{1}{2}$  approximation algorithm for this problem is presented in \cite{motwani}.

{\bf Our approach and results.} We start with computing a maximum weight {\em cycle cover} $C_{max}$  of $G$. A cycle cover of a graph $G$ is  a collection of cycles such  that each vertex belongs to exactly one of them. The weight of a maximum weight cycle cover $C_{max}$ is an upper bound on $opt$, where by $opt$ we denote the weight of a maximum weight traveling salesman tour.  By computing a maximum weight perfect matching $M$ we get another, even simpler than $C_{max}$,  upper bound -- on $opt/2$. From $C_{max}$ and $M$ we build a multigraph $G_1$ which consists of two copies of $C_{max}$ and one copy of $M$, i.e.,  for each edge $e$ of $G$  the multigraph $G_1$ contains between zero and three copies of $e$. Thus the total weight of the edges of $G_1$ is at least $\frac 52 \ opt$.  Next we would like to  {\em path-$3$-color} $G_1$, that is  to color the edges of $G_1$ with three colors, so that each color class
contains only vertex-disjoint paths. The paths from  the color class with maximum weight can then be patched in an arbitrary manner  into a tour of weight at least $\frac 56 \ opt$.

{\em Technique of eliminating difficult subgraphs via half-edges.} \  However, not every multigraph $G_1$ can be path-3-colored. For example, a subgraph of $G_1$ obtained  from a triangle $t$   of $C_{max}$  such that $M$  contains one of the edges of $t$ (such triangle is called a {\em $3$-kite (of $G_1$)}) cannot be path-3-colored as, clearly,  it is impossible to color such seven edges  with three colors and not create a monochromatic triangle.
Similarly, a subgraph of $G_1$ obtained from a square $s$  (i.e., a cycle of length four) of $C_{max}$  such that $M$  contains two edges connecting vertices of $s$  (such square is called a {\em $4$-kite (of $G_1$)}) is not path-3-colorable. To find a way around this difficulty, we compute another cycle cover $C_2$ {\em
improving $C_{max}$ with respect to $M$}, which is a cycle cover that does not contain any $3$-kite or $4$-kite of $G_1$ and whose weight is also at least $opt$.  An important feature of $C_2$ is that it may contain
{\em half-edges}. A half-edge of an edge $e$ is, informally speaking, a half of the edge $e$ that contains exactly one of its endpoints. Half-edges have already been introduced in \cite{PEZ}. Computing $C_2$ is done via a novel reduction to a maximum weight perfect matching. It is, to some degree, similar to computing a directed cycle cover without  length two cycles in \cite{PEZ}, but for Max TSP we need much more complex gadgets. 

 From one copy of $C_2$ and $M$ we build another multigraph $G_2$ with weight at least $\frac 32 \ opt$. It turns out that $G_2$ can always be {\em path-$2$-colored}. The multigraph $G_1$ may be non-path-$3$-colorable - if it contains at least one kite.  We notice, however, that if we remove one arbitrary edge from each kite, then $G_1$ becomes path-$3$-colorable.  The edges removed from $G_1$ are added to $G_2$. As a result,
the modified $G_2$ may stop being path-$2$-colorable. To remedy this, we in turn remove some edges from $G_2$ and add them to $G_1$.  In other words, we find  two disjoint sets of edges - a set $F_1 \subseteq G_1$ and a set $F_2 \subseteq G_2$, called {\em exchange sets} such that the multigraph $G'_1 = G_1 \setminus F_1 \cup F_2$ is path-$3$-colorable and the multigraph $G'_2 = G_2 \setminus F_2 \cup F_1$ is path-$2$-colorable. Since $G_1$ and $G_2$ have the total weight at least $4\  opt$, by path-$3$-coloring $G'_1$ and path-$2$-coloring $G'_2$ we obtain a $\frac 45$ - approximate solution to Max TSP.

{\em Edge coloring.} The presented algorithms for path-$3$-coloring and path-$2$-coloring are essentially based on a simple notion of a {\em safe edge}, i.e., 
an edge colored in such a way that it is guaranteed not to belong to any monochromatic cycle, used in an inductive way.  The adopted approach may appear simple and straightforward. For comparison, let us point out that the method of path-$3$-coloring the multigraph obtained from two directed cycle covers described in \cite{Svir} is rather convoluted.

Generally, the new techniques are somewhat similar to the ones used for the directed version of the problem - Max ATSP in \cite{Pal34}.
We are convinced that they will prove useful for other problems related with TSP, cycle covers or matchings.

The main result of the paper is
\begin{theorem}
There exists a $\frac 45$-approximation algorithm for Max TSP. Its running time is $O(n^3)$.
\end{theorem}

\section{Path-$3$-coloring of $G_1$}
We  compute a maximum weight cycle cover $C_{max}$ of a given complete undirected graph $G=(V,E)$ and a maximum weight perfect matching $M$ of $G$.
We are going to call cycles of length $i$, i.e., consisting of $i$ edges {\bf \em $i$-cycles}. Also sometimes $3$-cycles will be called {\bf \em triangles} and $4$-cycles -- {\bf \em squares}.
The multigraph $G_1$ consists of two copies of $C_{max}$ and one copy of $M$. We want to color each edge of $G_1$ with one of three colors of $\Kt =\{1,2,3\}$  so that each color class consists of vertex-disjoint  paths. 
The {\em graph} $G_1$ is a subgraph of the {\em multigraph} $G_1$ that contains an edge $(u,v)$  iff the multigraph $G_1$ contains an edge between $u$ and $v$. The path-$3$-coloring of $G_1$ can be equivalently defined as coloring each edge of (the graph) $G_1$ with the number of colors equal to the number of copies contained in the multigraph $G_1$. From this time on, unless stated otherwise, $G_1$ denotes a graph
and not a multigraph.

We say that a colored edge $e$ of  $G_1$  is {\bf \em safe}  if no matter how we color the so far uncolored edges of $G_1$ $e$ is guaranteed not to belong to any monochromatic cycle of $G_1$. An edge $e$ of $M$ is said to be {\bf \em external} if its two endpoints belong to two different cycles of $C_{max}$. Otherwise, $e$ is {\bf \em internal}.
We say that an edge $e$ is incident to a cycle $c$ if it is incident to at least one vertex of $c$.

We prove the following useful lemma.
\begin{lemma} \label{col}
Consider a partial coloring of $G_1$. Let $c$ be any cycle of $C_{max}$ such that for each color $k \in \Kt$ there exists an edge of $M$ incident to $c$ that is colored $k$. Then we can color $c$ so that each edge of $c$ and each edge incident to one of the edges of $c$ is safe.
\end{lemma}

\dowod   \ \ The proposed procedure of coloring $c$ is as follows.

If there exists an edge of $c$ that also belongs to $M$, we color it with all three colors of $\Kt$. For each uncolored edge of $M$ incident to $c$, we color it with an arbitrary color of $\Kt$.
Next, we orient the edges of $c$ (in any of the two ways) so that $c$ becomes a directed cycle $c$. Let $e=(u,v)$ be any  uncolored edge of $c$ oriented from $u$ to $v$. Then, there exists an edge $e'$ of $M$ incident to $u$. If $e'$ is contained in $c$, then we color $e$ with any two colors of $\Kt$. Otherwise $e'$ is colored with some color $k$ of $\Kt$.
Then we color $e$ with the two colors belonging to $\Kt \setminus k$.

First, no vertex of $c$ has three incident edges colored with the same color, as for each vertex its outgoing edge is colored with different colors than an incident matching edge.  Second, as for each color $k \in \Kt$ there is a matching edge incident to $c$ colored with $k$, there exists an edge of $c$ that is not colored $k$, thus $c$ does not belong to any color class, i.e. there exists no color $k \in \Kt$ such that each edge of $c$ is colored with $k$.  Let us consider now any edge $e=(u,v)$  of $M$ incident to some edge of $c$ and not belonging to $c$.  The edge $e$ is colored with some color $k$. 
Suppose also that  vertex  $u$ belongs to $c$ ($v$ may  belong to $c$ or  may not belong to $c$.) Let $u'$ be any other vertex of $c$ such that some edge of $M\setminus C_{max}$  colored $k$ is incident to it  ($u'$ may be equal to $v$ if $e$ is internal). 
To show that $e$ is safe, it suffices to show that there exists no path consisting of edges of $c \cup M$ that connects $u$ and $u'$ and whose every edge is colored $k$.  However, by the way we color edges of $c$ we know that the outgoing edges of $u$ and $u'$ are not colored with $k$ because of the way we oriented the cycle, there is no path connecting $u$ and $u'$ contained in $c$ that starts and ends with incoming edge. \koniec

For each cycle $c$ of $C_{max}$ we define its {\bf \em degree of flexibility} denoted as $flex(c)$ and its {\bf \em colorfulness}, denoted as $col(c)$. The degree of flexibility of a cycle $c$  is the number of internal edges of $M$ incident to  $c$
and the colorfulness of $c$ is the number of colors of $\Kt$ that are used for coloring the external edges of $M$ incident to $c$.

From Lemma \ref{col} we can easily derive 
\begin{lemma}\label{cola}
If a cycle $c$ of $C_{max}$ is such that $flex(c)+col(c) \geq 3$, then we can color $c$ so that each edge of $c$ and each edge incident to one of the edges of $c$ is safe.
\end{lemma}

Sometimes, even if a cycle $c$ of $C_{max}$ is such that  $flex(c)+col(c) < 3$, we can color the edges of $c$ so that each of them is safe.
For example, suppose that $c$ is a square consisting of edges $e_1, \ldots, e_4$  and there are four external edges of $M$ incident to $c$, all colored  $1$. Suppose also that each external edge incident to $c$ is already safe. Then we can color $e_1$ with $1$ and $2$,
$e_3$ with $1$ and $3$ and both $e_2$ and $e_4$ with $2$ and $3$. We can notice that $e_1$ is guaranteed not to belong to a cycle colored $1$ because external edges incident to $e_1$ are colored $1$ and are safe. Analogously, we can easily check that each other  edge of $c$ is safe. However, for example, a triangle $t$ of $C_{max}$ that has three external edges of $M$ incident to it, all colored with the same color of $\Kt$, cannot be colored in such a way that it does not contain a monochromatic cycle.

Consider a cycle $c$ of $C_{max}$ such that every external edge of $M$ incident to $c$ is colored. We say that $c$  is {\bf \em  non-blocked} if and only if (1)  $flex(c)+col(c) \geq 3$   or   (2)  $c$  contains at least $3 - flex(c)-col(c)$ vertex-disjoint edges, each of which has the property that it has exactly two incident external edges of $M$ and the two external edges of $M$  incident to it  are colored with the same color of $\Kt$ or (3) $c$ is a square such that $flex(c)=1$.

Otherwise we say that $c$ is {\bf \em blocked}.
We can see  that a cycle $c$ of $C_{max}$ is  blocked  if
\begin{itemize}
\item $c$ is a triangle and all external edges of $M$ incident to $c$ are colored with the same color of $\Kt$,
\item $c$ is a square with two internal edges of $M$ incident to it $(flex(c)=2)$,
\item $c$ is a cycle of even length, $flex(c)=0$   and  there exist two colors $k_1, k_2 \in \Kt$ such that external edges of $M$ incident to $c$ are colored alternately with $k_1$ and $k_2$.
\end{itemize}

Among blocked cycles we distinguish kites. We say that a cycle $c$ is a {\bf \em kite} if it is a triangle such that $flex(c)=1$ and then we call it a {\bf \em $3$-kite} or it is a square such that $flex(c)=2$ - called a {\bf \em $4$-kite}.
A cycle of $C_{max}$ which is not a kite is called {\bf \em unproblematic}.

Now, we are ready to present the algorithm for path-$3$-coloring $G_1$. \\

\begin{algorithm}
	\caption{Color $G_1$}
	\label{alg:col_g1}
	\begin{algorithmic}
	  \While{$\exists$ an uncolored external edge $e$ of $M$}
	    \State $c$ $\gets$ an unproblematic uncolored cycle of $C_{max}$ with the fewest 
	    uncolored external edges incident to $e$

	    \State color uncolored external edges incident to $c$ so that no unproblematic cycle
	    of $C_{max}$ becomes blocked and if possible, 
			
			\State so that $flex(c) + col(c) \geq 3$

	    \State color $c$ using Lemma \ref{coluzup} and internal edges incident to it in such a way, that
	    each edge incident to $c$ is safe
	  \EndWhile

	  \While{$\exists c$ -- an unproblematic, uncolored cycle of $C_{max}$}
	    \State color $c$ and internal edges incident to it in such a way, that
	    each edge incident to $c$ is safe
	  \EndWhile
	\end{algorithmic}
\end{algorithm}

\begin{lemma} 
Let $c$ be an unproblematic  cycle of $C_{max}$ that at some step of Algorithm Color $G_1$ has  the fewest uncolored external edges  incident to it. Then, it is always possible to color all uncolored external edges incident to $c$ so that
no unproblematic cycle of $C_{max}$ becomes blocked. Moreover, if   $c$ has at least two uncolored ext. edges incident to $c$  then, additionally, it is always possible to do it in such a way that  $flex(c)+col(c) \geq 3$.
If $c$ has exactly one  uncolored external edge $e$  of $M$ incident to it, then we can color $e$ so that  $flex(c)+col(c) \geq 3$ or so that $e$ is safe.
\end{lemma}
\dowod If $c$ has at least two uncolored external edges of $M$ incident to it, then we can use at least two different colors for coloring the edges. Moreover if $flex(c)=0$, then we can choose them in such a way that $col(c)=3$, i.e. so that for every color $k \in\Kt$ at least one external edge of $M$ incident to $c$ is colored with $k$. At this stage, every other uncolored cycle $c'$ of $C_{max}$ has also at least two uncolored external edges of $M$ incident to it. Therefore $c'$ is in danger of becoming blocked only if it has an even number of incident external edges of $M$,
 all of them are colored with the same two colors, say $k_1$ and $k_2$,  in an alternate way and it has exactly two incident uncolored external edges $e_1$, $e_2$  of $M$.  However,  even if we would like to also use $k_1$ and $k_2$ for coloring the external edges of $M$ incident to $c$, we can do it in such a way that $c'$ does not become blocked, because, as one can easily see, one of the ways of coloring $e_1$ and $e_2$ with $k_1$ and $k_2$
does not make $c'$ blocked.

If $c$ has exactly one uncolored external edge $e$ of $M$ incident to it and $c$ is in  danger of becoming blocked, then either $c$ is a triangle whose two other incident external edges are colored with the same color of $\Kt$  or  $c$ has even length and all of its  incident external edges of $M$  are colored with the same two colors in an alternate way. In each of these cases we have a choice  and  can color $e$ with one of two colors so that $c$ does not become blocked. If $e$ is  incident to a cycle $c'$ that is also in danger of becoming blocked,
then with respect to $c'$ we can also color $e$ with one of two colors of $\Kt$  so that it does not become blocked.  As the intersection of two two-element subsets of $\Kt$ is always nonempty, we can color $e$, say with $k$,  so that no cycle of $C_{max}$ becomes blocked. As all other external edges of $c$ were safe, then $e$ is also safe. \koniec

From the above lemma we get
\begin{corollary}\label{colsafe}
After all external edges are colored, each of them is incident to a cycle $c$ of $C_{max}$ such that  $flex(c)+col(c) \geq 3$ or is safe.
\end{corollary}

We say that a cycle $c$ of $G_1$ is {\bf \em  a subcycle}  of cycle $c'$ of $C_{max}$ if it goes only through vertices that belong to $c'$.

\begin{lemma} \label{coluzup}
Let $c$ be an unproblematic  and non-blocked cycle of $C_{max}$ whose all incident external edges of $M$ are already colored and safe.  Then it is always possible to color $c$ and internal  edges incident to $c$  in such a way that each edge incident to $c$ is safe.
\end{lemma} 

\dowod
If $c$ is such that   $flex(c)+col(c) \geq 3$, then by Lemmas \ref{col}  and \ref{cola},  the claim holds.
Now let us first  prove that if $c$ is not blocked and   $flex(c)+col(c) < 3$, then it is always  possible to color the edges of $c$ so that no color class contains all edges of  any subcycle of  $c$.

{\bf \em Case 1:}  All  edges of $M$ incident to $c$ are colored with the same color, say $k$. \\
We can  then assume that all edges of $M$ incident to $c$ are external. (Otherwise we would have colored internal edges  of $M$  with a different color than $k$.)
$c$ must have length at least $4$. (Otherwise it would be blocked.)  Let $k_1, k_2$  denote the two colors of $\Kt \setminus k$.
We choose two nonadjacent edges of $c$, color one of them with $k$ and $k_1$ and the other with $k$ and $k_2$. The remaining edges of $c$ are colored with $k_1$ and $k_2$.

{\bf \em Case 2:}  All  edges of $M$ incident to $c$ are colored with two colors, say $k_1$ and $k_2$. \\
We can assume that either (1)  $c$ has no incident internal edges of $M$  or (2)  that it has exactly one incident internal edge of $M$ and all external edges of $M$ incident to $c$ are colored in the same way.

Let $k$ denote the color belonging to $\Kt \setminus \{k_1, k_2\}$  and assume that $c$ goes through vertices $v_1, v_2, \ldots, v_s$ in the given order.
Then let $v_i$ denote a vertex of $c$ such that edges of $M$ incident to  $v_{i-1}$ and $v_i$ are colored in the same way, say with $k_1$, and $v_{i+1}$ is colored with $k_2$. Then for each $j \neq i$  we color  edge $(v_j, v_{j+1})$  of $c$ with colors belonging to 
$\Kt \setminus k'$, where $k'$ denotes the color used on an edge of $M$ incident to $v_j$. Edge $(v_i, v_{i+1})$ is going to be colored with $k_1$ and $k_2$. 

We colored the edges, so that there is no monochromatic cycle on edges of $c$ and internal matching edges. Therefore, together with the safety of all external edges, it ensures the safety of all internal edges. \koniec

\section{A cycle cover improving $C_{max}$ with respect to $M$}

Since $C_{max}$ may contain kites, we may not be able to path-$3$-color $G_1$. Therefore, our next aim is to compute another cycle cover  $C_2$ of $G$ such that it does not contain any cycle of $C_{max}$ which is problematic and whose weight is an upper bound on $OPT$. Since computing such $C_2$ may be hard, we relax the notion of a cycle cover and allow $C_2$ to contain {\bf \em half-edges}. A half-edge of the edge $e$ is informally speaking
a half of the edge $e$ that contains exactly one of the endpoints of $e$. Let us also point out  here that $C_2$ may contain kites which do not belong to $C_{max}$.
To be able to  give a formal definition of such a relaxed cycle cover, we introduce a graph $\tilde G$. We say that an edge $(u,v)$  is {\bf \em problematic} if $u$ and $v$ belong to the same kite. An edge connecting vertices of a kite $c$  is also said to be a problematic edge of $c$. A $3$-kite has no diagonals  and a $4$-kite has two diagonals.   $\tilde G=(\tilde V, \tilde E)$ is the graph obtained from $G$ by splitting each problematic  edge $(u,v)$ with a vertex $x_{\{u,v\}}$ into two edges 
$(u,  x_{\{u,v\}})$ and $(x_{\{u,v\}}, v)$, each with weight $\frac 12w(u,v)$.  Each of the edges $(u,  x_{\{u,v\}})$ and $(x_{\{u,v\}}, v)$ of $\tilde G$ is said to be {\bf \em a half-edge of the edge $(u,v)$ of $G$}.  In what follows, when we speak of an edge of a kite, we mean an edge of the original graph $G$.

\begin{definition}\label{relst}

A {\bf \em relaxed cycle cover   improving $C_{max}$ with respect to $M$}  is a subset $\tilde C\subseteq \tilde E$ such that
\begin{itemize}
\item[(i)]
each vertex in $V$ has exactly two incident edges  in $\tilde C$;

\item[(ii)]
for each $3$-kite  $t$ of $C_{max}$  the number of  half-edges  of  the edges of $t$ contained in  $\tilde C$  is even  and   not greater than four;
\item[(iii)]
for each $4$-kite  $s$ of $C_{max}$ the number of half-edges  of  the edges  or diagonals of $s$  contained in $\tilde C$ is even  and not greater than six.
\end{itemize}
\end{definition}

To compute a relaxed cycle cover  $C_2$  improving $C_{max}$  with respect to $M$  we construct the following  graph $G'=(V',E')$.
The set of vertices $V'$  is a superset of the set of verices $V$ of $G$.
For each problematic edge $(u,v)$  of $G$ we add two vertices $x_{\{u,v\}}^u, x_{\{u,v\}}^v$ to $V'$ and edges $(u, x_{\{u,v\}}^u), (x_{\{u,v\}}^v,v)$ to $E'$. For each problematic edge $(u,v)$ which is not a diagonal of a $4$-kite we add also an edge $(x_{\{u,v\}}^u, x_{\{u,v\}}^v)$. The edge $ (x_{\{u,v\}}^u, x_{\{u,v\}}^v)$ has weight $0$ in $G'$
and each of  the edges   $(u, x_{\{u,v\}}^u),  (x_{\{u,v\}}^v,v)$  has weight  $\frac{1}{2}w(u,v)$.
Each of the vertices  $x_{\{u,v\}}^u, x_{\{u,v\}}^v$  is called {\bf \em a splitting vertex of the edge $(u,v)$}.
For each edge $(u,v)$ of $G$ which is not problematic we add an edge $(u,v)$ to $E'$ with weight $w(u,v)$. 

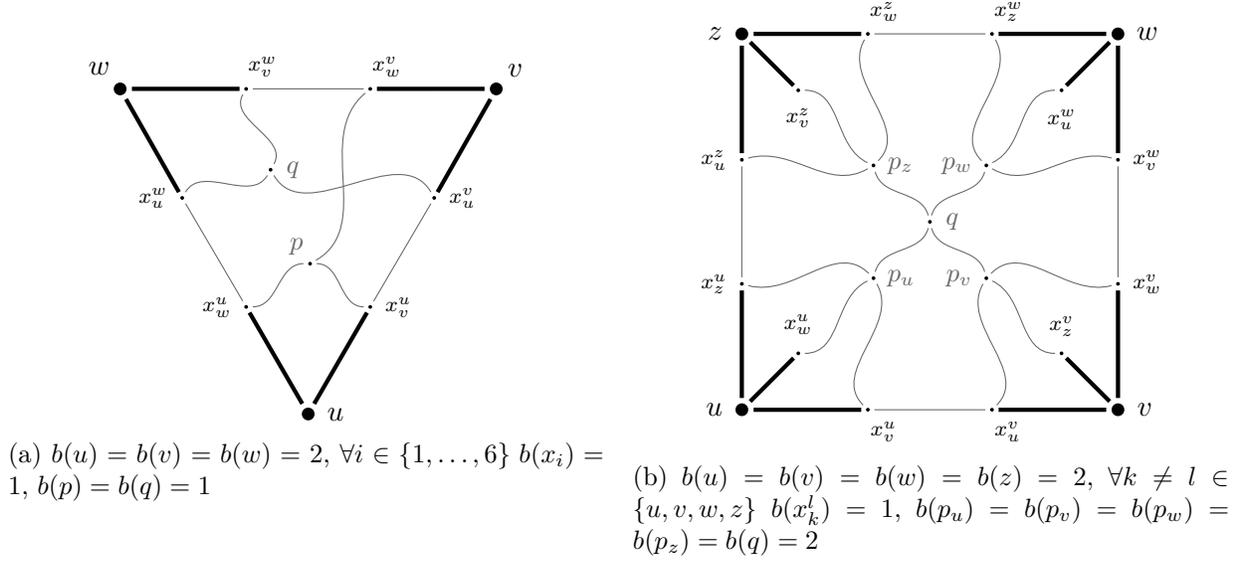
\begin{figure}[h!]
    
    \begin{subfigure}{.48\textwidth}
  \centering
  \begin{tikzpicture}[scale=2.5]
    \tikzset{every node/.style={draw,fill=black,circle,scale=1,inner sep=0pt,outer sep=2pt}}
    \tikzstyle{pom}=[draw=black!60,thin]
    \node[vert,label={right:$u$}] (a) at (1,0) {};
    \node[vert,label={45:$v$}] (b) at (2,1.73) {};
    \node[vert,label={135:$w$}] (c) at (0,1.73) {};

    \node[label={right:\scriptsize{{$x_v^u$}}}] (aab) at (1.33,.57) {};
    \node[label={right:\scriptsize{{$x_u^v$}}}] (abb) at (1.67, 1.15) {};

    \node[label={85:\scriptsize{{$x_w^v$}}}] (bbc) at (1.33,1.73) {};
    \node[label={85:\scriptsize{{$x_v^w$}}}] (bcc) at (.67,1.73) {};

    \node[label={left:\scriptsize{{$x_u^w$}}}] (acc) at (.33, 1.15) {};
    \node[label={left:\scriptsize{{$x_w^u$}}}] (aac) at (.67, .57) {};

    \draw[sol] (a) -- (aab);
    \draw[sol] (abb) -- (b);
    \draw[sol] (b) -- (bbc);
    \draw[sol] (bcc) -- (c);
    \draw[sol] (c) -- (acc);
    \draw[sol] (aac) -- (a);

    \draw[pom] (aab) -- (abb);
    \draw[pom] (bbc) -- (bcc);
    \draw[pom] (acc) -- (aac);

    \node[label={110:\textcolor{black!60}{\small{$p$}}}] (Pa) at (1.01, .8) {};
    \draw[pom] (aab) to [out=180, in=0] (Pa);
    \draw[pom] (aac) to [out=0, in=180] (Pa);
    \draw[pom] (bbc) to [out=220, in=30] (Pa);

    \node[label={right:\textcolor{black!60}{\small{$q$}}}] (Pc) at (.8,1.3) {};
    \draw[pom] (bcc) to [out=240, in=60] (Pc);
    \draw[pom] (acc) to [out=60, in=240] (Pc);
    \draw[pom] (abb) to [out=120, in=300] (Pc);
  \end{tikzpicture}
  \caption{$b(u) = b(v) = b(w) = 2$, $\forall i \in \{1,\dots,6\}\; b(x_i) = 1$,
  $b(p) = b(q) = 1$}
\end{subfigure}
~
\begin{subfigure}{.48\textwidth}
  \centering
  \begin{tikzpicture}[scale=2.5]
    \tikzset{every node/.style={draw,fill=black,circle,scale=1,inner sep=0pt,outer sep=2pt}}
    \tikzstyle{pom}=[draw=black!60,thin]
    \node[vert,label={left:$u$}]   (a) at (-1, -1) {};
    \node[vert,label={right:$v$}]   (b) at (1, -1) {};
    \node[vert,label={right:$w$}]   (c) at (1, 1) {};
    \node[vert,label={left:$z$}]   (d) at (-1, 1) {};

    \node[label={275:\scriptsize{{$x_v^u$}}}] (aab) at (-.33,-1) {};
    \node[label={275:\scriptsize{{$x_u^v$}}}] (abb) at (.33, -1) {};

    \node[label={right:\scriptsize{{$x_w^v$}}}] (bbc) at (1, -.33) {};
    \node[label={right:\scriptsize{{$x_v^w$}}}] (bcc) at (1, .33) {};

    \node[label={85:\scriptsize{{$x_z^w$}}}] (ccd) at (.33, 1) {};
    \node[label={85:\scriptsize{{$x_w^z$}}}] (cdd) at (-.33, 1) {};

    \node[label={left:\scriptsize{{$x_u^z$}}}] (add) at (-1, .33) {};
    \node[label={left:\scriptsize{{$x_z^u$}}}] (aad) at (-1, -.33) {};

    \node[label={90:\scriptsize{{$x_w^u$}}}] (aac) at (-.7, -.7) {};
    \node[label={below:\scriptsize{{$x_u^w$}}}] (acc) at (.7, .7) {};
    \node[label={above:\scriptsize{{$x_z^v$}}}] (bbd) at (.7, -.7) {};
    \node[label={below:\scriptsize{{$x_v^z$}}}] (bdd) at (-.7, .7) {};

    \draw[sol] (a) -- (aab);
    \draw[sol] (abb) -- (b);
    \draw[sol] (b) -- (bbc);
    \draw[sol] (bcc) -- (c);
    \draw[sol] (c) -- (ccd);
    \draw[sol] (cdd) -- (d);
    \draw[sol] (d) -- (add);
    \draw[sol] (aad) -- (a);

    \draw[pom] (aab) -- (abb);
    \draw[pom] (bbc) -- (bcc);
    \draw[pom] (ccd) -- (cdd);
    \draw[pom] (add) -- (aad);

    \draw[sol] (a) -- (aac);
    \draw[sol] (b) -- (bbd);
    \draw[sol] (c) -- (acc);
    \draw[sol] (d) -- (bdd);

    \node[label={right:{\textcolor{black!60}{\small{$p_u$}}}}] (pu) at (-.3, -.3) {};
    \draw[pom] (aab) to [out=135, in=300] (pu);
    \draw[pom] (aac) to [out=0, in=205] (pu);
    \draw[pom] (aad) to [out=345, in=135] (pu);

    \node[label={left:{\textcolor{black!60}{\small{$p_v$}}}}] (pv) at (.3, -.3) {};
    \draw[pom] (abb) to [out=45, in=240] (pv);
    \draw[pom] (bbd) to [out=180, in=335] (pv);
    \draw[pom] (bbc) to [out=195, in=45] (pv);

    \node[label={left:{\textcolor{black!60}{\small{$p_w$}}}}] (pw) at (.3, .3) {};
    \draw[pom] (bcc) to [out=165, in=330] (pw);
    \draw[pom] (acc) to [out=180, in=25] (pw);
    \draw[pom] (ccd) to [out=300, in=135] (pw);

    \node[label={right:{\textcolor{black!60}{\small{$p_z$}}}}] (pz) at (-.3, .3) {};
    \draw[pom] (add) to [out=15, in=210] (pz);
    \draw[pom] (bdd) to [out=0, in=155] (pz);
    \draw[pom] (cdd) to [out=240, in=45] (pz);

    \node[label={right:{\textcolor{black!60}{\small{$q$}}}}] (q) at(0,0) {};
    \draw[pom] (pu) to [out=75, in=255] (q);
    \draw[pom] (pv) to [out=105, in=285] (q);
    \draw[pom] (pw) to [out=255, in=75] (q);
    \draw[pom] (pz) to [out=285, in=105] (q);
  \end{tikzpicture}
  \caption{$b(u) = b(v) = b(w) = b(z) = 2$, $\forall k \neq l \in \{u,v,w,z\}\;
  b(x_k^l) = 1$, $b(p_u) = b(p_v) = b(p_w) = b(p_z) = b(q) = 2$}
\end{subfigure}

    \caption{Gadgets for 3-kites~\textbf{(a)} and
    4-kites~\textbf{(b)} of~$G_1$ in graph~$G$. Half-edges corresponding to the
    original edges are thickened, the auxiliary edges are thin. Original
    vertices (thick dot) are connected with all the other original vertices of
    graph~$G$. The auxiliary vertices have no connections outside of the gadget.
    The figures are subtitled with the specifications of $b(v)$ values for
    different vertices. For a vertex $t$ with $b(t) = i$, the resulting
    b-matching will contain exactly $i$ edges ending in $t$.}
    \label{fig:maxtsp_gadgets}
\end{figure}

Next we build so-called gadgets.
For each $3$-kite $t$ on vertices $u,v,w$ we add two vertices $p^t, q^t$ to $V'$. Let's assume that $u$ is incident to external edge of $M$.  Vertex $p^t$ is connected to the splitting vertices of edges of $t$ that are neighbors of $u$, i.e. to vertices $x_{\{u,v\}}^u, x_{\{u,w\}}^u$ and to vertex $x_{\{v,w\}}^v$.  Vertex $q^t$ is connected to every other splitting vertex of $t$, i.e. $x_{\{u,w\}}^w, x_{\{v,w\}}^w, x_{\{u,v\}}^v$. All edges incident to verices $p^t, q^t$
have weight $0$ in $G'$.

For each $4$-kite of $C_{max}$  on vertices $u,v,w,z$    we add five vertices  $p^s_u, p^s_v, p^s_w, p^s_z, q^s$ to $V'$.  Vertex $p^s_u$ is connected to the splitting vertices of edges of $s$ that are neighbors of $u$, i.e. to vertices $x_{\{u,v\}}^u, x_{\{u,w\}}^u, x_{\{u,z\}}^u$.  Vertices $p^s_v, p^s_w, p^s_z$ are connected analogously. Vertex $q$ is connected to vertices $p^s_u, p^s_v, p^s_w, p^s_z$. All edges incident to verices $p^s_u, p^s_v, p^s_w, p^s_z, q^s$
have weight $0$.

We will reduce the problem of computing a relaxed cycle cover improving $C_{max}$ with respect to $M$ to the problem of computing a perfect $b$-matching of the graph $G'$.
We define the function $b: V' \rightarrow {\cal N}$ in the following way. For each vertex $v \in V$ we set $b(v)=2$. For each splitting vertex $v'$ of some problematic edge we set $b(v')=1$.
For all  vertices $p^t$ and $q^t$, where $t$ denotes a $3$-kite of $C_{max}$ we have $b(p^t)=b(q^t)=1$.
For all  vertices $p^s_u$ and $q^s$, where $s$ denotes a $4$-kite of $C_{max}$ and $u$ one of its vertices we have $b(p^s_u)=b(q^s)=2$.

\begin{theorem}
Any perfect $b$-matching of $G'$ yields  a relaxed cycle cover  $C_2$  improving $C_{max}$ with respect to $M$.
A maximum weight perfect $b$- matching of $G'$ yields a relaxed cycle cover $C_2$ improving $C_{max}$  with respect to $M$ such that $w(C_2) \geq OPT$.
\end{theorem}
\dowod  First we  show that any perfect $b$-matching of $G'$ yields  a relaxed cycle cover  improving $C_{max}$ with respect to $M$.
Let $B$ be any perfect $b$-matching of $G'$. $B$ defines  $C_2 \subseteq \tilde E$ as follows. A half-edge $(u, x_{\{u,v\}})$  is contained in $C_2$ iff edge  $(u, x^u_{\{u,v\}})$ of $G'$ is contained in $B$. A non-problematic edge $(u,v)$ is contained in $C_2$ iff $(u,v)$ is contained in $B$.
Since $b(v)=2$ for any vertex $v$ of $V$, we can see that  the property $(i)$ of Definition \ref{relst} is satisfied.

Consider now an arbitrary $3$-kite $t$ of $C_{max}$. There are 3 problematic edges of $t$ and thus six half-edges. Suppose that $t$ is on vertices $u,v,w$. We can notice that a half-edge $(u, x_{\{u,v\}})$  is not contained in $C_2$ 
iff a splitting vertex $x^u_{\{u,v\}}$ is connected in $B$ to one of the vertices $p^t, q^t$ or to a splitting vertex $x^v_{\{u,v\}}$. Since $p^t$ and $q^t$ are connected to one splitting vertex each, at most 4 half-edges of the problematic edges of $t$ are contained in $B$. If
 a splitting vertex $x^u_{\{u,v\}}$ is connected in $B$ to  $x^v_{\{u,v\}}$, then both half-edges of the edge $(u,v)$ are excluded from $C_2$.
This shows that the number of half-edges of problematic edges of $t$ contained in $C_2$ is even.

Consider now an arbitrary problematic square $s$ of $C_{max}$. There are six problematic edges of $s$ and thus twelve half-edges of these edges. Suppose that $s$ is on vertices $u,v,w,z$. We can notice that a half-edge $(u, x_{\{u,v\}})$  is not contained in $C_2$ 
iff a splitting vertex $x^u_{\{u,v\}}$ is not matched to $u$ in $B$. Thus a half-edge $(u, x_{\{u,v\}})$ does not occur in $C_2$ iff 
 a splitting vertex $x^u_{\{u,v\}}$ is connected in $B$ to one of the vertices $p^s_u, p^s_v, p^s_w, p^s_z$ or to a splitting vertex $x^v_{\{u,v\}}$.
Since $q$ is connected to two of the vertices $p^s_u, p^s_v, p^s_w, p^s_z$ and $b(p^s_u)=b(p^s_v)= b(p^s_w)= b(p^s_z)=2$, exactly six splitting vertices of the problematic edges of $s$ are connected in $B$ to vertices $p^s_u, p^s_v, p^s_w, p^s_z$. This means  that at least six half-edges of the problematic edges of $s$ are not contained in $B$.  If
 a splitting vertex $x^u_{\{u,v\}}$ is connected in $B$ to  $x^v_{\{u,v\}}$, then both half-edges of the edge $(u,v)$ are excluded from $C_2$.
This shows that the number of half-edges of problematic edges of $s$ contained in $C_2$ is even.

In order to show that $w(C_2) \geq OPT$ it suffices to prove the following lemma.
\begin{lemma}
\label{lem:cycle_cover_optimality}
Every cycle cover not containing kites of $G_1$ corresponds to some  perfect b-matching of $G'$.
\end{lemma} 
The proof is in Section \ref{sec:c2_opt_proof} \koniec

\section{Exchange sets $F_1, F_2$ and path-$2$-coloring of $G'_2$}

We construct a multigraph $G_2$ from one copy of  a relaxed cycle cover $C_2$ and one copy of a maximum weight perfect matching $M$.
Since $C_2$ may contain half-edges and we want $G_2$ to  contain only edges of $G$, for each half-edge of edge $(u,v)$ contained in $C_2$, we will either include the whole edge $(u,v)$ in $G_2$ or not include it at all. While doing so we have to ensure that the total weight of the constructed multigraph $G_2$ is at least $\frac 32 opt$. 

 The main idea behind deciding which half-edges are extended to full edges and included in $G_2$ is that we compute two sets $Z_1$ and  $Z_2$ such that for each kite in $G_1$ half of the edges containing half-edges belongs to $Z_1$ and the other half to $Z_2$. (Note that by Lemma \ref{} each kite in $G_1$ contains an even number of half-edges in $C_2$.) Let $I(C_2)$ denote the set consisting of whole edges of $G$ contained in $C_2$. This way $w(C_2)= w(I(C_2)) +\frac 12 (w(Z_1)+w(Z_2))$. Next, let $Z$ denote the one of the sets $Z_1$ and $Z_2$ with maximum  weight. Then $G_2$ is defined as a multiset consisting of edges of $M$, edges of $I(C_2)$ and edges of $Z$. We obtain

\begin{fact}
The total weight of the constructed multigraph $G_2$ is at least $\frac 32 opt$.
\end{fact} 
\dowod
The weight of $M$ is at least $\frac 12 opt$. The weight of $w(C_2)= w(I(C_2)) +\frac 12 (w(Z_1)+w(Z_2))$ is at least $opt$. Since $w(Z) = \max\{w(Z_1), w(Z_2)\}$, we obtain
that $w(I(C_2))+ w(Z) \geq w(C_2)$. \koniec

Since $C_{max}$ contains at least one kite, $G_1$ is non-path-$3$-colorable. We can notice, however, that if we remove one edge from each kite from the multigraph $G_1$, then the obtained multigraph is path-$3$-colorable.

If we manage to construct a set $F_1$ with one edge per each kite such that additionally the multigraph $G_2 \cup F_1$ is path-$2$-colorable, then we have a $\frac 45$-approximation of Max TSP. Since computing such $F_1$ may be difficult, we allow, in turn, certain edges of $C_2$ to be removed from $G_2$ and added to $G_1$. Thus, roughly,  our goal is to compute such disjoint sets $F_1, F_2$ that:

\begin{enumerate}

\item $F_1 \subset C_{max}$ contains at least  one edge of each kite;
\item for each kite $c$, $F_2 \subset I(C_2)$ contains exactly one edge not contained in $c$; 
\item the multigraph $G'_1=G_1 \setminus F_1 \cup F_2$ is path-$3$-colorable;
\item the multigraph $G'_2=G_2 \setminus F_2 \cup F_1$ is path-$2$-colorable.
\end{enumerate}

Let $F_1$ and $F_2$ be two sets of edges that satisfy properties 1. and 2. of the above. Then  the set of edges $C'_2=(I(C_2) \cup Z \cup F_1) \setminus F_2$  can be partitioned into {\bf \em cycles and paths of $G'_2$}, where $G'_2$ denotes the resulting multigraph $G_2 \setminus F_2 \cup F_1$.  The partition of $C'_2$ into cycles and paths is carried out in such a way that two incident edges of $C'_2$ belonging to a common path or cycle of $C_2$, belong also to a common path or cycle of $C'_2$ (and $G'_2$). Also, the partition is maximal, i.e., we cannot
add any edge $e$ of $C'_2$ to any path $p$ of $G'_2$ so that $p \cup \{e\}$ is also a path or cycle of $G'_2$.

We say that $e$ is a {\bf \em double edge} of $G'_2$, or that $e$ is {\bf \em double}, if the multigraph $G'_2$ contains two copies of $e$. In any path-$2$-coloring of $G'_2$ every double edge must have both colors of $\Kd$ assigned to it. 

We observe that in order for $G'_2$ to be path-$2$-colorable,  we have to guarantee that  there does not exist   a cycle
$c$  of $G'_2$ of odd length $l$ that  has $l$ incident double edges. Since every two consecutive edges of $c$ are incident to some double edge, they must be assigned different colors of $\Kd$ and because the length of $c$ is odd, this is clearly impossible. The way to avoid 
this is to choose one edge of each such potential cycle and add it to $F_2$. 

We say that a path $p$ of $G'_2$ beginning at $w$ and ending at  $v$  is {\bf \em amenable} if (i) neither $v$ nor $w$ has degree $4$ in $G'_2$
or (ii) $v$ has degree $4$, $w$ has degree smaller than $4$ and $p$ ends with a double edge, the last-but-one edge of $p$ is a double edge or the last-but-one and the last-but-three vertices in $p$ are matched in $M$.

It turns out that $G'_2$ that does not contain odd cycles described above and whose every path is amenable is path-$2$-colorable - we show it in
 Section \ref{path2}. To facilitate the construction of $G'_2$, whose every path is amenable and to ensure that $F_1$ and $F_2$ have certain other useful properties we create  two opposite orientations of $I(C_2)$: $D_2$ and $opp(D_2)$. In each of these orientations $I(C_2)$ contains directed cycles and paths and each kite has the same number of incoming and outgoing edges. (This can be achieved by pairing the endpoints of paths ending at the same kite and combining them. For example, if  $C_2$ contains half-edges $h_1=(u, x^u_{\{u,v\}})$ and $h_2=(w, x^w_{\{u,w\}})$ of a certain $3$-kite $t$ and edges $e_1=(u',u), e_2=(w',w)$, then in the orientation  in which $e_1$ is directed from $u'$  to $u$ the edge $e_2$ must be directed from $w'$ to $w$.)  Apart from whole edges $C_2$ contains also half-edges. Let $H(C_2)$ denote the set of edges of $G$ such that $C_2$ contains exactly
one half-edge of each of these edges. We partition $H(C_2)$ into two sets $Z_1, Z_2$ so that for each kite $c$ half of the edges of $H(C_2)$
is contained in $Z_1$ and the other half in $Z_2$.
With each of the orientations $D_2, opp(D_2)$ we associate
one of the sets $Z_1, Z_2$. Thus,  we assume that $D_2$ contains $Z_1$, with the edges of $Z_1$ being oriented in a consistent way with the edges of $I(C_2)$ under orientation $D_2$, and $opp(D_2)$ contains $Z_2$,  with its edges being oriented accordingly.
The exact details of the construction of $Z_1$ and $Z_2$ are given in the proof of Lemma \ref{F12}.

Depending on which of the sets $Z_1, Z_2$ has bigger weight,  we either choose the orientation $D_2$ or $opp(D_2)$.  Hence, from now on, we assume that the edges of $I(C_2) \cup Z$ are directed.

\begin{lemma}\label{F12}
It is possible to compute sets  $F_1, F_2$   such that they and  the resulting $G'_2$ satisfy:
\begin{enumerate}
\item $F_1 \subset C_{max} \setminus ((Z \cup I(C_2))\cap M)$;
\item $F_2 \subseteq I(C_2) \cup Z$;
\item for each kite $c$, (i) the set $F_1$ contains exactly one edge of $c$ and the set $F_2$ contains zero edges of $c$ or (ii) (it can happen only for $4$-kites) the set $F_1$ contains exactly two edges of $c$ and the set $F_2$ contains one edge of $c \setminus M$;
\item for each kite $c$ the set $F_2$ contains exactly one outgoing edge of $c$;
\item for each kite $c$ and each vertex $v$ of $c$ the number of edges of $F_2$ incident to $v$ is at most one greater than the number of edges of $F_1$ incident to $v$;
\item there exists no cycle of $G'_2$ of odd length $l$ that has $l$ double edges incident to it;
\item each path of $G'_2$ is amenable.

\end{enumerate}
\end{lemma}

The property 1. of this lemma guarantees that $G'_2$ does not contain more than two copies of any edge. We  show in Appendix \ref{path2} that properties 6. and 7. are essentially sufficient for the multigraph $G'_2$ to be path-$2$-colorable. Properties  4. and 5. will be helpful
in finding a path-$3$-coloring of $G'_1$. Property 5. ensures that no vertex $v$ has six incident edges in $G'_1$. 

The proof of this lemma is given in Section \ref{dowodf12}.

The path-$2$-coloring of $G'_2$ is quite similar to the path-$3$-coloring of $G_1$. It is described in Section \ref{path2}.

\section{Completing the path-coloring of $G$}
After the construction and path-$2$-coloring of $G'_2$ we are presented with the task of extending the partial path-$3$-coloring of $G_1$ to the complete path-$3$-coloring of $G'_1$. In particular, we have to color the  edges
of kites, edges of $F_2$ that have been added during the construction of $G'_2$ and external edges of $M$ incident to $3$-kites, called {\bf \em tails}. A tail incident to a $3$-kite $t$ is said to be a {\bf \em tail of $t$}.

Let us now describe the set of uncolored edges of $G'_1$ in more detail. Each one of them is incident to some kite and has either (1) two endpoints belonging to the same kite $c$ (an internal edge of $c$), or (2) one of its endpoints belongs to some kite $c$ and the other does not belong to any kite (an external edge of $c$) or (3) its endpoints belong to two different kites $c$ and $c'$ (an external edge both of $c$ and $c'$). Let $t$ denote a $3$-kite. Then by Lemma \ref{F12} exactly one edge  of $t$ belongs to $F_1$, no edge of $t$ belongs to $F_2$ and there also exists  exactly one edge in $e \in F_2$  that is an outgoing edge of $t$, i.e., $e$ is an external edge of $t$ and is directed from an endpoint belonging to $t$ in $I(C_2)$.
$F_2$ may also contain up to three incoming edges of $t$, each one incident to a different vertex of $t$.  Any incoming edge of $t$ is also an outgoing edge of some other kite. A tail of $t$ is also uncolored in $G'_1$. 

Each uncolored edge $e$ of $G'_1$ has a requirement $d(e)$ denoting the number of colors of $\Kt$ that must be assigned to it. 
 Then for any edge $e$ contained in some $3$-kite, $d(e)=3$ if $e \in M \setminus F_1$ , $d(e)=1$ if $e \in F_1\setminus M$ and $d(e)=2$ otherwise. Thus, for each $3$-kite $t$ we have to color exactly six of its edges in the {\em multigraph} $G'_1$.

Let $s$ denote a $4$-kite. Then by Lemma \ref{F12} either (1) exactly one edge  of $s$ belongs to $F_1$ and no edge of $s$ belongs to $F_2$ or (2) exactly two edges of $s$ belong to $F_1$ and one edge of $s$ belongs to $F_2$. There also exists  exactly one edge $e \in F_2$  that is an outgoing edge of $s$. $F_2$ may also contain up to four incoming edges of $s$, each one incident to a different vertex of $s$. For any edge $e$ belonging to some $4$-kite, $d(e)=3$ if $e \in M \setminus F_1$ or $e \in F_2$, $d(e)=1$ if $e \in F_1\setminus M$ and $d(e)=2$ otherwise.  Thus, for each $4$-kite $s$ we have to color exactly nine of its edges in the {\em multigraph} $G'_1$.

Each uncolored external edge $e$ in $G'_1$ has requirement $d(e)=1$. 
Let $H$ denote the subgraph of $G'_1$ comprising all edges  with positive requirement.

We need to assign colors of $\Kt$ to edges of $H$ (or, in other words, color edges of $H$ with colors of $\Kt$) in such a way that each color class in the whole graph $G'_1$  forms a collection of disjoint paths. The coloring of edges of $H$ is an extension of the already existing partial path-$3$-coloring of $G_1$. Therefore, for some edges there exist restrictions on colors of $\Kt$ that can be assigned to them. Consider any vertex $v$ that does not belong to any kite and that has one or two incident edges in $H$. If $v$ has an incident tail in $H$, then it has exactly two incident edges in $G'_1\setminus H$   that are colored with two different pairs of colors of $\Kt$ (while path-$3$-coloring $G_1$ we can easily guarantee that two consecutive edges of $C_{max}$ incident to $v$ such that an edge of $M$ incident to $v$ is also incident to a $3$-kite are colored with two different pairs of colors). Let these pairs of colors be $\{k_1, k_2\}$ and $\{k_2, k_3\}$. Hence any edge of $H$ incident to $v$ may be colored only with $k_1$ or $k_3$ - we associate with $v$  a  two-element subset $Z(v)=\{k_1, k_3\}$.  Moreover, if  $v$ has two incident edges in $H$ and we color one of them with $k_1$, then the other one {\em must} be colored with $k_3$.  If $v$ does not have an incident tail in $H$, then it has at most one incident edge in $H$ and exactly five edges in the {\em multigraph} $G_1$ as well as in the multigraph  $G'_1\setminus H$. In this case there exists exactly one color $k$  of $\Kt$ that can be assigned to an edge of $H$ incident to $v$ and we associate a one-element subset $Z(v)=\{k\}$ with $v$.

Let $t$ be a $3$-kite. Then a vertex of $t$ incident to its tail  is called a {\bf \em foot vertex (of $t$)}. If $e' \in F_1 \cap t$ is incident to the foot vertex of $t$, then $t$ is said to be {\bf \em vertical}; otherwise it is {\bf \em horizontal}.
Two $3$-kites $t_1$ and $t_2$ having a common tail are called {\bf \em twins}. Also, each one of them is called a twin and $t_1$ is said to be a {\bf \em brother of $t_2$}. A $3$-kite that is not a twin is said to be {\bf \em non-twin}.

Some of the edges contained in $H$ are directed.  The directions of edges of $H$ satisfy:

\begin{enumerate}
\item each internal edge is undirected (i.e., each edge contained in a kite);
\item the direction of each edge of $F_2$ is the same as in $I(C_2)$; the properties of edges of $F_2$ are described in Lemma \ref{F12} in properties (3), (4) and (5);

\item a tail of  two twins is undirected; otherwise, a tail of a $3$-kite $t$ is an incoming edge of $t$. (It may happen that a tail $e$ of some $3$-kite belongs also to $I(C_2)$ and $F_2$. Then $G'_1$ contains two copies of $e$, each one with the requirement $d(e)=1$ and the copy corresponding to a tail is treated as a tail and the other copy is treated as an external directed edge.)
\end{enumerate}

From graph $H$ we build a graph $I$ by shrinking each kite to a single vertex. Each vertex of $I$ that corresponds to a kite in $H$ is called, respectively a {\bf \em t-vertex} (if it is a $3$-kite) or an {\bf \em s-vertex} (if it is a $4$-kite); each remaining vertex is called an {\bf \em o-vertex}.  In any coloring of $I$ or $H$, we say that an o-vertex $v$ is {\bf \em respected} if any edge incident to  $v$ is assigned a color belonging to $Z(v)$ and if there are two edges incident to $v$, then they have different colors assigned to them.

To {\bf \em pre-color} a  directed cycle  or path $r$ of $I$ means to
color each of its edges with a color of $\Kt$ so that each o-vertex of $r$ is respected.  To {\bf \em color a kite $c$} means to  color each edge $e$ of $c$ with $d(e)$ colors of $\Kt$. 

We are going to color the edges of $H$ in portions - by considering directed cycles and paths in $I$. For each such cycle or path we will color its edges as well as some of the kites corresponding to its vertices. To be able to talk more precisely about these operations we introduce below the notions of {\bf \em processing} a directed cycle or path $r$ in $I$ and {\bf \em step-processing} a vertex $v$ on $r$.
Processing a directed cycle or path $r$ in $I$ consists in step-processing each of its vertices on $r$.

\begin{definition}
Let $r$ be a directed cycle or path in $I$ and
 $v$ a vertex on $r$  that has an outgoing edge that belongs to $r$.

To {\em step-process} $v$ (or in case $v$ corresponds to a kite $c$, to step-process $c$) {\em on $r$} means: 
\begin{itemize}
\item if an outgoing edge of $v$ is uncolored - to color it,
\item if $v$ has  an incoming edge contained in $r$ - to color it,
\item if $v$ corresponds to a kite $c$ -  to color the kite $c$ unless $c$ is a horizontal twin, whose brother has not been step-processed (on any directed cycle or path in $I$),
\item if $v$ corresponds to a non-twin $3$-kite $t$ -  to color the tail of $t$,
\item if  $v$ corresponds to a twin $3$-kite $t$, whose brother $t'$ has already been step-processed - to color the common tail of $t$ and $t'$ and in case $t'$ has not already been colored, to color $t'$,
\item to carry out the above so that each color class forms a collection of vertex-disjoint paths in $G'_1$ and so that each o-vertex in $I$ is respected.
\end{itemize}
\end{definition}

To {\bf \em process} a directed  path $r$ in $I$ that goes through vertices $v_1,  \ldots, v_k$ and directed from $v_k$ to $v_1$ means to step-process each of the vertices $v_2, \ldots, v_k$ in turn, starting from $v_2$.
When we process such a path, then we start the step-processing $v_2$ by coloring an outgoing edge of $v_2$ incident also to $v_1$.   We then continue step-processing  $v_2$ and afterwards, proceed to ste-processing $v_3$, then $v_4$ and so on. If $v_i$ and $v_j$ of $r$ correspond to twins $t_i$ and $t_j$ such that $t_i$ is horizontal and considered before $t_j$ on $r$, then while step-processing $t_i$ we only color the edges incident to $v_i$ and leave $t_i$ and its tail uncolored. When we come to $v_j$, we color the incoming edge of $r$ incident to $v_j$ and both twins $t_i$ and $t_j$ and their common tail.
In an analogous way we define the processing of a directed cycle $r$ in $I$ - we start from  step-processing any vertex on $r$ and continue with step-processing subsequent vertices along $r$.

Let us notice that if a vertex $v$ corresponding to a kite $c$ has not been step-processed, then $c$ is uncolored and either (1) every external 
edge of $c$ is also uncolored or (2) an outgoing edge of $c$ is colored because we have just step-processed $v'$ on some directed path or cycle $r$ such that $r$ contains an edge $(v,v')$; apart from this every other external edge of $c$ is uncolored. Also, a given vertex $v$  has exactly one outgoing edge in $I$ but may belong to more than one directed path in $I$ or it may belong to a directed cycle and some directed path(s) in $I$.  However, in Algorithm 2  the first time we encounter $v$ while processing a directed cycle or path, we will step-process it, because each considered directed path is maximal under inclusion. If we encounter $v$ again
while processing a different cycle or path, we will just color some of its incoming edges (and possibly a tail and so on) but will not step-process $v$ again.

\begin{algorithm} \label{AH}
	\caption{Color $H$}
	\label{alg:col_h}
	\begin{algorithmic}
	  \While{$\exists$ a directed cycle in $I$}
	    \State process it and remove its edges from $I$
	  \EndWhile
		\While{$\exists$ a directed maximal path in $I$}
	    \State process it and remove its edges from $I$
	  \EndWhile
		
	\end{algorithmic}
\end{algorithm}

\vspace{0.5cm}

In Section \ref{AHcor} we  prove that every  directed cycle or path can be processed.

\section{Summary}

\noindent \fbox{
\begin{minipage}[t]{\textwidth}
\vspace{0.5cm}
{\bf \em \hspace{0.5cm} Algorithm MaxTSP}
\vspace{0.5cm}
\begin{enumerate}
\item Compute a cycle cover $C_{max}$ of $G$ of maximum weight and a perfect matching $M$ of $G$ of maximum weight.
\item Let $G_1$ denote a multigraph obtained from two copies of $C_{max}$ and one copy of $M$ - its weight is at least $\frac 52 opt$.
Path-$3$-color $G_1$ with colors of $\Kt=\{1,2,3\}$ leaving kites and edges of $M$ incident to kites uncolored.
\item Compute a maximum weight relaxed cycle cover $C_2$ improving $C_{max}$ with respect to $M$.
\item  Let $G_2$ denote a multigraph obtained from one copy of $C_2$ and one copy of $M$ - its weight is at least $\frac 32 opt$. Compute the sets of edges $F_1 \subset C_{max}, \ F_2 \subset C_2$ such that the multigraph $G'_1=G_1 \setminus F_1 \cup F_2$ is path-$3$-colorable
and  the multigraph $G'_2=G_2 \setminus F_2 \cup F_1$ is path-$2$-colorable (i.e. $F_1, F_2$ are as in Lemma \ref{F12}).
\item Path-$2$-color $G'_2$ with colors of $\Kd=\{4,5\}$.
\item Extend the partial path-$3$-coloring of $G_1$ to the complete path-$3$-coloring of $G'_1$.
\item Choose the color class of maximum weight - its weight is at least $\frac 45 opt$ and complete the paths of this class into a traveling salesman tour in an arbitrary way.
\end{enumerate}
\vspace{0.5cm}
\end{minipage}
}

\vspace{1cm}

The presented algorithm works for graphs with an even number of vertices. If the number of vertices of a given graph is odd, then we can guess one edge, shrink it and compute the remaining part of the solution in the graph with even vertices. 

\section{Correctness of Algorithm 2} \label{AHcor}

We  are going to prove that every  directed cycle or path in $I$ can be processed.
First we give several auxiliary lemmas.

\begin{lemma}\label{preparz}
Let $c$ be a  directed cycle in $I$ of even length,   whose every other vertex is an o-vertex. Then we are able to pre-color  $c$ in such a way that its every two consecutive edges get assigned different colors.
\end{lemma}
\dowod

First, let us notice that  an o-vertex $v$ may have two incident edges in $I$ only if one of them is a tail of some $3$-kite. Thus, every vertex  of $c$ that is not an o-vertex must correspond to a $3$-kite and be a t-vertex.

If the length of $c$ is two, then $c$ contains exactly one o-vertex $v$. We then assign one  color of $Z(v)$ to one edge of $c$ and the other color of $Z(v)$ to the other edge of $c$ and are done. 

Suppose now that $c$ has length greater than two. Let $v$ be any o-vertex of $c$ and $e_1, e_2$ the edges of $I$ incident
to $v$. We assign one of the colors $k_1$ of $Z(v)$ to $e_1$ and the other $k_2$ to $e_2$. Assume that $e_2$ is an incoming edge of $v_1$, $e_3$ is an outgoing edge of $v_1$  and $e_4, \ldots, e_k$ are the subsequent edges of $c$. The edges $e_3$ and $e_4$ are incident
 to another o-vertex $v'$ of $c$. We will show now that whatever the set $Z(v')$, we are always able to asign colors to $e_3$ and $e_4$ in such a way that $e_4$ {\em does not} get assigned $k_1$ - the color already assigned to $e_1$. If $Z(v')$ contains $k_2$ and some other color $k_4$, then we assign $k_4$ to $e_3$ and $k_2$ to $e_4$. If $Z(v')$ does not contain $k_2$, then it contains $k_1$ and $k_3$ and we assign $k_1$ to $e_3$ and $k_3$ to $e_4$. This way (i)  edges $e_2$ and $e_3$  get assigned  different colors
and (ii) $e_4$ gets assigned a color different from $k_1$. 

If $c$ has length $4$, then we notice that the edges $e_1$ and $e_4$  of $c$ get assigned different colors as well and we are done.

If $c$ has length greater than $4$, then we consider the next pairs of edges and continue in the manner described above.  More precisely, when we consider the pair of edges $e_{2i+1}$ and $e_{2i+2}$ incident to some o-vertex $w$,
we know that the invariant that $e_1$ and $e_{2i}$ have different colors assigned is satisfied. Our goal is to color $e_{2i+1}$ and $e_{2i+2}$ in such a way that (i) $e_{2i+1}$ gets assigned a color different from the color assigned to $e_{2i}$ and (ii) $e_{2i+2}$ gets assigned a color different from $k_1$. From the way we have analysed coloring $e_3$ and $e_4$, we know that it can always be done.
 
\koniec 

\begin{corollary} \label{cyklparz}
Let $c$ be a  directed cycle in $I$ of even length,   whose every other vertex is an o-vertex. Then we are able to process  $c$.
\end{corollary}
\dowod First, let us notice that every $t$-vertex of $c$ corresponds to a non-twin $3$-kite, because the tail of each such kite is contained in $c$ and thus is directed. 

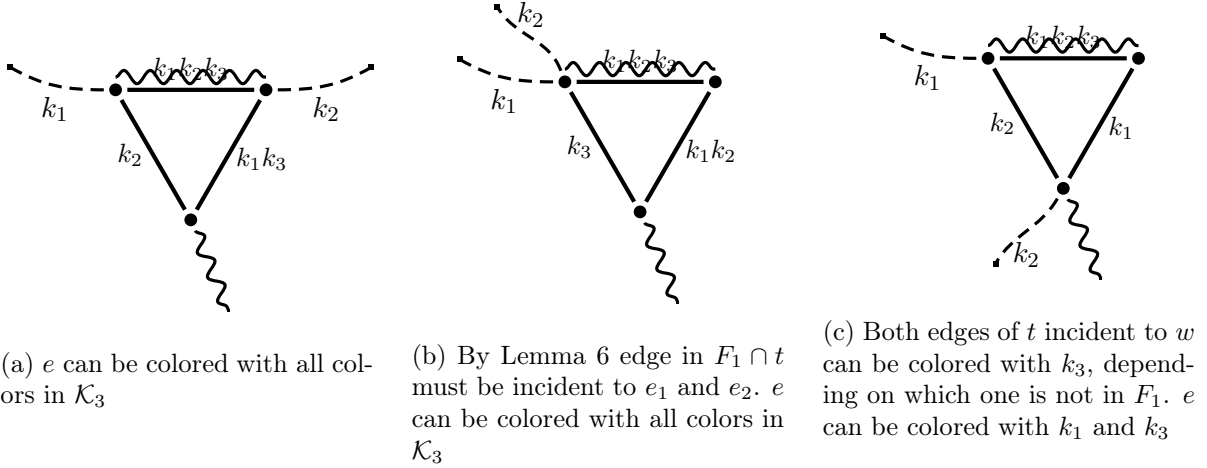
\begin{figure}[h!]
\centering
	\begin{subfigure}{0.3\textwidth}
	\centering
		\begin{tikzpicture}
			\trikite{}{}
			\tlExta{$k_1$}{}
			\trExta{$k_2$}{}
			\tLabel{$k_1 k_2 k_3$}
			\lLabel{$k_2$}
			\rLabel{$k_1 k_3$}
		\end{tikzpicture}
		\subcaption{$e$ can be colored with all colors in $\Kt$}
	\end{subfigure}
	\quad
	\begin{subfigure}{0.3\textwidth}
	\centering
		\begin{tikzpicture}
			\trikite{}{}
			\tlExta{$k_1$}{}
			\tlExtb{$k_2$}{}
			\tLabel{$k_1 k_2 k_3$}
			\lLabel{$k_3$}
			\rLabel{$k_1 k_2$}
		\end{tikzpicture}
		\subcaption{By Lemma \ref{F12} edge in $F_1 \cap t$ must be incident to $e_1$ and $e_2$. $e$ can be colored with all colors in $\Kt$}
	\end{subfigure}
	\quad
	\begin{subfigure}{.3\textwidth}
		\centering
		\begin{tikzpicture}
			\trikite{}{}
			\tlExta{$k_1$}{}
			\blExtb{$k_2$}{}
			\tLabel{$k_1 k_2 k_3$}
			\lLabel{$k_2$}
			\rLabel{$k_1$}
		\end{tikzpicture}
		\subcaption{Both edges of $t$ incident to $w$ can be colored with $k_3$, depending on which one is not in $F_1$. $e$ can be colored with $k_1$ and $k_3$}
	\end{subfigure}
	\caption{Vertical triangles with $k_1\neq k_2$ and $e_1$ not incident to $w$}
	\label{fig:lemma_compl}
\end{figure}

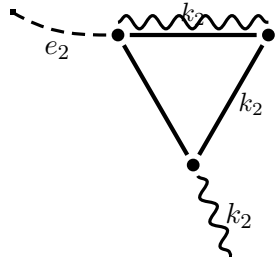
\begin{figure}[h!] \label{fighoriz}
	\centering
	\begin{tikzpicture}
		\trikite{$k_2$}{}
		\tlExta{$e_2$}{}

		\rLabel{$k_2$}
		\tLabel{$k_2$}
	\end{tikzpicture}

\end{figure}

While pre-coloring $c$ whenever two edges $e_1, e_2$ of $c$ adjacent to the same kite $c$ get colored, we also color $c$.
While coloring $c$ we only have to see to it that no vertex of $c$ gets three incident edges of the same color in $G'_1$ and to that $c$ does not contain a monochromatic cycle i.e. a $3$-cycle. We show how to color $c$ in Figure \ref{fig:lemma_compl} and in Figure \ref{fighoriz}.
 Let us notice that after pre-coloring 
$c$ and all $3$-kites corresponding to t-vertices on $c$, no color class contains a cycle - this is because every edge $e$ of $c$ is incident to a t-vertex corresponding to a $3$-kite $t$ and the only external edges incident to $t$ in the whole graph $G'_1$ are $e$ and some other edge $e'$ of $c$. We know, however, that every two consecutive edges of $c$ are colored differently. Hence $e'$ is colored differently from $e$. Thus, neither $e$ nor $e'$ can belong to a monochromatic cycle, which means that  in this way we process $c$.
\koniec

Suppose that the tail $e$ of $t$ is uncolored. Then $t$ is said to be {\bf \em flexible} if there exist such two colors $k,k' \in \Kt$ that $e$ can be colored both with  $k$ and $k'$, by which we mean that if we color the tail of $t$ with $k$ (or correspondingly $k'$), then the foot of $t$ does not have more than three incident edges colored with $k$ (resp. $k'$). The flexibility of a $3$-kite $t$ is useful when $t$ is a vertical twin that is step-processed before its twin $t'$. Then while step-processing $t$ we color $t$ but leave its tail uncolored and later later while step-processing $t'$ we have a greater 'flexibility' in coloring $t'$ and its tail.

\begin{lemma}\label{compl}
Let $t$ be any uncolored vertical $3$-kite and $e_1, e_2 \in F_2 \setminus M$ two external edges incident to $t$ colored with, respectively, $k_1$ and $k_2$. Let $w$ be the foot vertex  and $e$ the tail of $t$. Additionally, $e_1$ and $e_2$ are not both incident to $w$ and  $k_1 \neq k_2$. Then it is possible to color the edges of $t$ so that $t$ becomes flexible and so that $e$ can be colored with $k_3 \notin \{k_1, k_2\}$. 
\end{lemma}
\dowod For all possible triangles we will show how to color the edges for $t$. These colorings are presented in Figure \ref{fig:lemma_compl}.

\koniec

\begin{lemma} \label{wlasciwosci}
The computed sets $F_1, F_2$ satisfy:
\begin{enumerate}
\item No foot of a $3$-kite has two incident edges of $F_2$.
\item If a $3$-kite  has four incident edges of $F_2$, then it is vertical.
\end{enumerate}
\end{lemma}
The proof follows from the proof of Lemma \ref{F12}.

\begin{lemma} \label{ver1}
Let $t$ be a vertical $3$-kite, whose tail $e$ is uncolored and that has been colored at some point as in Lemma \ref{compl}. Then, however, we color any further  external edges of $H$ incident to $t$ apart from its tail, $t$ always stays flexible.
\end{lemma}
\dowod The lemma follows from the fact that the foot of $t$ has not two incident edges of $F_2$. \koniec
 
Let $t$ be an uncolored $3$-kite $t$, whose tail $e$ is also uncolored. Then we say that $t$ is {\bf \em weakly flexible} if  there exist two colors $k, k' \in \Kt $ such that $t$ can be colored in at least two ways and
in one of these colorings $e$ can be colored with $k$ and in the other with $k'$, i.e., after coloring $e$ with $k$ or $k'$, the foot of $t$ has at most two incident edges colored with respectively $k$ or $k'$.
We say that an uncolored twin $t$ is {\bf \em versatile} if every two colored edges of $H$ incident to $t$ have different colors assigned to them. The weak flexibility of a $3$-kite $t$ is useful when $t$ is a horizontal twin that is step-processed before its twin $t'$.  While step-processing $t$ on some directed cycle or path $r$ we do not color it or its tail but only the incident edges of $r$ and later while step-processing $t'$ we color both $t$ and $t'$ and their common tail.

\begin{lemma} \label{horiz}
Every uncolored versatile horizontal $3$-kite is weakly flexible.
\end{lemma}
\dowod Let $t$ be any triangle on vertices $u,v,w$ as in Lemma \ref{horiz} and let $e_1, e_2, e_3$ three external edges incident to $t$ colored with, respectively, $k_1$, $k_2$ and $k_3$. Let $w$ be the foot vertex and $e$ the tail of $t$. Let us assume that $e_2$ and $e_3$ are not incident to $w$. Then we can color $e$ with $k_2$ and $k_3$. For each of these colors we have to show how to color edges of $t$. As these cases are symmetric, we assume that $e$ is colored with $k_2$. Let us assume that $e_2$ is incident to $v$. Then we color $(u,v)$ and $(u,w)$ with $k_2$.

As $t$ is horizontal, we still have to color $(v,w)$ with $2$ colors, and the other edges with one color. If there is an edge, say $e_1$, incident to $w$ we color $(v,w)$ and $(v,u)$ with $k_1$. If there is an  edge incident to $v$ other than $e_2$, say $e_3$, we color $(v,w)$ and $(u,w)$ with $k_3$. If there are both of these edges, than it is correct coloring. Otherwise there is an edge incident to $u$, say $e_1$, and we can color $(v,w)$ and either $(u,v)$ or $(u,w)$ with $k_1$, so we can always color $t$. \koniec

\begin{lemma} Every  directed cycle or path can be processed in such a way that at all times every uncolored horizontal twin is versatile.
\end{lemma} 
\dowod
Let us consider a directed path $p$ going through vertices $v_1, \ldots, v_k$ and directed from $v_k$ to $v_1$. 
We can notice that since cycles are processed before paths, each vertex of $p$ is distinct.
We observe also that $v_1$ is either an o-vertex or corresponds to a kite that has already been step-processed - otherwise we could extend $p$, because then the outgoing edge of $v_1$ would be uncolored.  Vertex $v_k$, on the other hand, is either an o-vertex or corresponds to an uncolored (and not step-processed) kite. We begin by  coloring the arc $(v_2, v_1)$ with any color of $\Kt$ that is available. 
Let us note that some color of $\Kt$ is always available because of the following.  If $v_1$ is an o-vertex, then it has exactly six incident edges in the {\em multigraph} $G'_1$ - apart from five edges in the {\em multigraph} $G_1$, it  has an additional incoming edge that is an outgoing edge of some kite. If $v_1$ corresponds to a kite, then  Lemma \ref{F12} Property 5 guarantees that any vertex in $G'_1$ belonging to a kite has degree at most six.

 Also, if $v_1$ corresponds to an uncolored $3$-kite $t$ that has already been step-processed, then we color $(v_2, v_1)$ with such a color $k$ of $\Kt$ that no external edge of $t$ is colored with $k$.  Such a color $k$ always exists because only horizontal $3$-kites can be left uncolored and they have at most three incident edges of $F_2$. Thus we can guarantee that $t$ remains versatile.

We step-process subsequent vertices on $p$ according to the rules listed below.

Let $e_1 \in p$ be an outgoing edge of $v$ colored with $k_1$ and $e_2$ an uncolored incoming edge of $v$. Depending on whether $v$ is an o-, t- or s-vertex and other conditions we proceed
as follows:
\begin{enumerate}
\item $e_2 \in M$. Then $v$ must be a t-vertex corresponding to a $3$-kite $t$ and $e_2$ is an outgoing edge of an o-vertex $v'$. We color $t$, $e_2$ and an incoming edge $e_3$ of $v'$. If $Z(v')=\{k_1, k'\}$, then we  color $e_2$ with $k'$ and $e_3$ with $k_1$. Otherwise $Z(v')=\{k_2, k_3\} = \Kt \setminus k_1$. Then we color $e_2$ with $k_2$ and $e_3$ with $k_3$ or the other way around.

\item $v$ is a t-vertex corresponding to a non-twin $3$-kite $t$ and $e$ is the tail of $t$ incident to an o-vertex $v'$. We color $t, e_2$ and $e$.  If $Z(v')=\{k_1, k'\}$, then we  color $e$ with $k'$ and $e_2$ with $k \in \Kt \setminus \{k_1, k'\}$. Otherwise $Z(v')=\{k_2, k_3\} = \Kt \setminus k_1$. Then we color $e_2$ with $k_2$ and $e$ with $k_3$ or the other way around.

\item $v$ is a t-vertex corresponding to a twin $3$-kite $t$, whose brother $t'$ has already been colored and $e$ is the tail of  $t$. Then by Lemmas \ref{compl} and \ref{ver1}, there exist two colors of $\Kt$, such that if we look only at $t'$, then $e$ can be colored with either of them. Let $Z(e)$ denote the set consisting of these two colors.  Now we proceed almost identically as in the case above. If $Z(e)=\{k_1, k'\}$, then we  color $e$ with $k'$ and $e_2$ with $k \in \Kt \setminus \{k_1, k'\}$. Otherwise $Z(e)=\{k_2, k_3\} = \Kt \setminus k_1$. Then we color $e_2$ with $k_2$ and $e$ with $k_3$ or the other way around. We color $t$. Note  that  each colored external edge of $t$ is colored with a different color. This cannot be said about $t'$ - it may happen that the tail of $t'$ is colored with the same color as some other external edge $e'$ of $t'$, but we do not have to worry about edge $e'$ ending in a monochromatic cycle because then the tail of $t$ and $t'$ would also have to belong to such cycle.

\item $v$ is a t-vertex corresponding to a twin $3$-kite $t$, whose brother $t'$ has not been step-processed.  We color $e_2$ with $k_2$ or $k_3$.
If $t$ is vertical, then we color $t$. Otherwise we leave $t$ uncolored.

\item $v$ is a t-vertex corresponding to a twin $3$-kite $t$, whose brother $t'$ has  been step-processed but is uncolored. Since $t'$ has been step-processed, but is uncolored, it is horizontal.  By Lemma \ref{horiz} $t'$ is weakly flexible - therefore there exist two colors $k, k'$ that can be used for coloring the tail of $t'$. If $k_1 \in \{k,k'\}$, then we color the tail of $t$ and $t'$ with the color belonging to  $\{k,k'\} \setminus k_1$ and $e_2$ with the remaining color of $\Kt$ - note that this way each colored external edge of $t$ is colored with a different color.  If $k_1 \notin \{k,k'\}$, then we color the tail with $k_2$ and $e_2$ with $k_3$.  We also color both $t$ and $t'$.

\item $v$ is an s-vertex. Then we color $e_2$ with any color of $\Kt$ different from $k_1$. We also color $s$.

\end{enumerate}

We argue that by proceeding as above, we do not create a monochromatic cycle in $G'_1$ and thus process $p$. This is so, because every external edge colored with $k$ while processing $p$ is contained in some  path $p'$ consisting of edges colored with $k$ that ends at a vertex of $p$ corresponding to a kite $c'$ such that $c'$ has only one incident external edge in $G'_1$ colored with $k$.

Let us now turn our attention to directed cycles. Let $c$ be a  directed cycle of $I$. We can assume that $c$ is not as in Lemma \ref{preparz}, because we have already dealt with such cycles. Thus $c$ contains two subsequent vertices  $v, v'$ such that neither $v$ nor $v'$ is an o-vertex and $c$ contains an edge $(v,v')$ directed from $v$ to $v'$. If possible we choose $v$ that is an s-vertex or corresponds to a $3$-kite $t$, whose tail also belongs to $c$. If such $v$ does not exist then we choose $v$ that corresponds to a $3$-kite that is either non-twin or whose brother has already been step-processed. For now, we assume that this is the case.

 We start by coloring an incoming edge of $v$. If $v$ corresponds to a $3$-kite $t$, whose tail $q$ does not belong to $c$, then $q$ can be colored with some two colors $k_2, k_3$. In this case we color the incoming edge of $v$ with $k_1 \notin \{k_2, k_3\}$.

We continue processing $c$ according to the rules described above until we reach the vertex $v'$. If applying the rules also to $v'$ would result in the arc $(v,v')$ being colored with a different color than an incoming edge of $v$, we apply the rules to $v'$ and are done. Otherwise $v'$ must correspond to a $3$-kite $t'$ which is either non-twin or that is a twin whose brother has already been colored. Also, assume  that the incoming edge of $v$ is colored with $k_1$. It follows  that the tail of $t'$ can be colored with $k_2$ or $k_3$, both different from $k_1$, and that the outgoing edge of $t'$ is colored with $k_2$ or $k_3$. Otherwise we would be able to color $(v,v')$ with a color different from $k_1$. Suppose that the outgoing edge of $v$ is colored with $k_2$. In this case we color the tail of $t'$ with $k_3$ and color $(v,v')$ with $k_3$ - if it is not incident to the tail of $t'$ and with $k_2$ otherwise. By Lemma \ref{wlasciwosci} property 1, it cannot happen that both the outgoing and incoming edge of $t'$ is incident to the foot of $t'$.

We are left with the case when each vertex of $c$ corresponds to a twin $3$-kite whose brother also occurs on $c$. We leave this case to the reader. \koniec

\section{Path-$2$-coloring} \label{path2}
The partition of $G'_2$ into cycles and paths is carried out in such a way that two edges of $C'_2$ belonging to a common path or cycle of $C_2$, belong also to a common path or cycle of $G'_2$. Also, the partition is maximal, i.e., we cannot
add any edge $e$ of $C'_2$ to any path $p$ of $G'_2$ so that $p \cup \{e\}$ is also a path or cycle of $G'_2$. We may assume that each path and cycle of $G'_2$ is directed - the orientations of edges are consistent with those in $G_2$.

A {\bf \em surrounding} of a cycle $c$ of $C_2$, denoted as $sur(c)$, contains  every edge of $c$ and  every edge of $M$ incident to $c$.
Let $p$ be a path of $G'_2$ directed from $u$ to $v$. If $v$ has degree $4$ in $G'_2$, then an edge $e$ of $p$ incident to $v$ is said to be a {\bf \em border} of $p$.
The {\bf \em surrounding} of  $p$, denoted as $sur(p)$, contains  every edge of $p$  and every edge of $M$ incident to $p$.  

We construct a directed graph $G_p= (V_p, E_p)$ such that each path of $G'_2$ is represented by some vertex of $V_p$ and $E_p$ contains an edge $(p, p')$ iff $p$ has a border and the border of $p$ is incident to some vertex of $p'$.
Thus each vertex of $V_p$ has at most one outgoing edge. Below we describe the algorithm for path-$2$-coloring the graph $G'_2$. In it we first color the cycles of $G'_2$ and their surroundings.  The order of coloring the paths of $G'_2$
is dictated by the structure of graph $G_p$: we begin by coloring the paths of $G'_2$ that form cycles in $G_p$; next at each step we color an uncolored path, whose outdegree in $G_p$ is zero.

The presence of borders complicates path-$2$-coloring in two aspects:
\begin{enumerate}
\item Suppose that edges $e_1=(u,u')$ and $e_2=(u,u'')$ belong to some path of $G'_2$ and that $u$ is incident to a double edge $e_d$ different from $e_1$ and $e_2$.
Since $e_d$ has to be colored with two colors of $\Kd$, edges $e_1$ and $e_2$ must be assigned different colors of $\Kd$. Therefore while path-$2$-coloring $G'_2$ we will preserve the following invariant:
\begin{invariant}\label{invdouble}
Every two edges $e_1=(u,u'), e_2=(u,u'')$ of $G'_2$ such that their common endpoint $u$ is incident to a double edge $e_d$ different from $e_1$ and $e_2$ are assigned different colors of $\Kd$. 
\end{invariant}

\item Each border $b$ of a path $p$ of $G'_2$ is colored while coloring the path $p$ and not before. In particular, if $b$ is double and is incident to a path or cycle $p'$ such that $p'$ is colored before $p$, then while coloring $p'$ we assign only one color to $b$. The second one is assigned while coloring $p$.
If $b$ is double we may also think of it as of two edges - one being a border and the other an edge of the matching $M$.

Because of this we modify the meaning of a safe edge in this section as follows. We say that a colored edge $e$ is safe if no matter how we color the so far uncolored edges except for any uncolored borders, $e$ is guaranteed not to belong to any monochromatic cycle. In particular, it means that if we want to prove that a newly colored border $b$ is safe we have to explicitly show that it does not belong to any monochromatic cycle - without taking use of the fact that previously colored 
edges are safe.

\end{enumerate}

\begin{algorithm}
  \caption{Color $G'_2$}
  \label{alg:col_gprim2}
  \begin{algorithmic}
	  \State During the whole execution ensure that Invariant \ref{invdouble} is satisfied. 
    \While{$\exists c$ -- an uncolored cycle of $G'_2$}
      \For{$e \in sur(c)$}
        \State color $e$ in such a way, that it is safe
      \EndFor
    \EndWhile 
    
    	\While{$\exists c_p$ -- a directed cycle of $G_p$}
      \For{$p$ such that $p$ is a vertex on $c_p$}
        \State color each $e \in sur(p)$ in such a way, that it is safe
      \EndFor
      \State remove each vertex of $c_p$ together with incident edges from $G_p$
    \EndWhile
    
    \While{$\exists p$ -- an uncolored path of $G'_2$ such that $outdeg_{G_p}(p)=0$}
      \For{$e$ such that $e \in sur(p)$}
        \State color  $e$ in such a way, that it is safe
      \EndFor
      \State remove $p$ together with incident edges from $G_p$
    \EndWhile

    \end{algorithmic}
\end{algorithm}

\begin{lemma} \label{cykl45}
Let $c$ be an uncolored cycle $c$ of $C_2$ considered at some step of Algorithm Color $G_2$. Then it is possible to color each edge belonging to $sur(c)$ in such a way that it is safe.
\end{lemma}
\dowod The procedure of coloring the edges of $sur(c)$ is similar to that described in the proofs of Lemmas \ref{col} and \ref{coluzup}. 
We orient the edges of $c$ so that $c$ becomes directed. 

{\em Case 1:} (i) For each color $k \in \Kd$ there exists an edge of $M$ incident to one of the edges of $c$ that is colored $k$  or (ii) there exists an uncolored edge of $M$ incident to one of the edges of $c$. 
First we color every uncolored non-double edge $e$ of $M$ incident to $c$ so that case (i) holds. Next we color each double edge incident to $c$.
Let $e=(u,v)$ be a double edge such that $u$ belongs to $c$. Then, necessarily  $v$ belongs to some path of $C_2$ and since we color cycles of $C_2$ before coloring paths of $C_2$, $e$ is uncolored. We start with such a double edge $e=(u,v)$ that the predecessor $u'$ of $u$ on the cycle $c$ has no incident double edge. The existence of such double edge is guaranteed by Lemma \ref{F12}. Let $e_1=(u,u')$ and $e_2=(u, u'')$ be two edges of $c$ incident to $u$ and let $e'$ be an edge of $M$ incident to $u'$ and $k$ the color of $\Kd$ assigned to $e'$.  To preserve Invariant \ref{invdouble} we  have to color the edges $e_1, e_2$ with different colors of $\Kd$. To make it possible we color $e$ with a color $k'$ belonging to $\Kd \setminus k$, i.e., for the time being we color $e$ only with one color instead of two.  
We proceed with each subsequent double edge incident to $c$ in the same way, i.e., we color such edges in order of their occurrence along $c$. 

Further we color all edges of $c$. Let $e=(u,v)$ be an edge of $c$ oriented from $u$ to $v$ and let $e'$ be an edge of $M$ incident to $u$
$k$ the color of $\Kd$ assigned to $e'$. Then we color $e$ with a color $k'$ belonging to $\Kd \setminus k$. We can notice that each so far colored edge is safe. Suppose that $e=(u,v)$ is colored with $k$. Then we additionally assign $k'\neq k$ to $e$.

{\em Case 2:} All edges of $M$ incident to $c$ are colored with the same color $k$. \\
We color any chosen one edge of $c$ with $k$ and the remaining ones with $k' \neq k$.  \koniec

\begin{lemma}\label{path45}
Let $p$ be an  uncolored path $p$ of $C_2$ considered at some step of Algorithm Color $G_2$ such that $outdeg_{G_p}(p)=0$. Then it is possible to color each edge belonging to $sur(p)$ in such a way that it is safe.
\end{lemma}

\dowod 
Generally  we proceed in a very similar way as in Lemma \ref{cykl45}. The path $p$ is already oriented.  First we color each edge  of $M$ incident to $p$ with one color of $\Kd$ in order of their occurrence along $p$. If a given edge $e$ of $M$ incident to $u$ is double, then  we color it with one color only and with the one different from that assigned to an edge of $M$ incident to $u'$ which proceeds $u$ on $p$. Next we color each edge $(u,v)$ of $p$ directed from $u$ to $v$, which is not a border of $p$ with a color different from that assigned to an edge $e'$ of $M$ incident to $u$. 

We must also color the border $b$ of $p$, if $p$ has one. 

If $b$ is double, then it must have got assigned one color of $\Kd$ before we started coloring $p$ - that is because $outdeg_{G_p}(p)=0$, which means that $b$ got colored while coloring the path or cycle of $G'_2$ incident to $b$. It may also happen that the border $b$ of $p$ is incident to some ''internal'' vertex of $p$ but then we have also already assigned one color of $\Kd$ to it. If $b$ is already colored with $k_1$,  then we additionally assign $k_2\neq k_1$ to it. 
The safety of $b$ follows from the following. The edge $e$ proceeding $b$ on $p$ is colored with one color $k$ of $\Kd$. From the way we color edges of $p$, we notice that $e$ is contained in a monochromatic path $p_k$ colored with $k$, whose one endpoint lies on $p$. In other words we claim that $p_k$ has a ''dead end''. We can observe  that a part of $p_k$ starting with $e$ is contained in $p$ and does not leave $p$. It follows from the fact that  each edge $(u,v)$ of $p$ is colored with a color different from the one assigned to the edge of $M$ incident to $u$. This means that $e$ is safe, because we have already colored every edge of $p$ and every edge of $M$ incident to $p$ (except possibly for some borders), hence $b$ is safe.

If the border $b=(u,v)$ of $p$ is not double, then we still have to color it. Suppose that $v$ is the endpoint of $p$. Then three edges of the multigraph $G'_2$ incident to $v$ have already been colored. This means that there is only one color of $\Kd$ that can be used for coloring $b$. We must also ensure that after coloring $b$, it does not belong to any monochromatic cycle. Since $p$ is amenable, $b$ is either proceeded by a double edge on $p$ or an edge $e'$ of $M$ incident to $v$ is also incident to a last-but-three vertex of $p$. In the first case, the safety of $b$ follows from the fact that an edge of $p$ proceeding a double edge proceeding $b$ is safe. (The argument is the same as above.) In the second case 
we leave the edge $e'$ uncolored till this point. Once we know that we are forced to color $b$, with say $k \in \Kd$, we color $e'$ with the other color of $\Kd$ and we also color accordingly the two edges proceeding $b$ and are done.
\koniec

\begin{figure}
  \centering
\begin{subfigure}[t]{0.47\textwidth}
	\centering
	\begin{tikzpicture}
		\node[vert,label={below right:$v_2$}] (a) at (0,0) {};
		\node[vert] (b) at ([shift={(a)}] 15:1.25) {};
		\node[vert,label={below right:$u_1$}] (d) at ([shift={(b)}] 45:1.25) {};
		\node[vert,label={above:$v_1$}] (e) at ([shift={(d)}] 75:1.25) {};
		\node[vert] (f) at ([shift={(e)}] 195:1.25) {};
		\node[vert,label={right:$u_2$}] (g) at ([shift={(f)}] 225:1.25) {};

		\node[vert,label={above:$w_1$}] (i) at ([shift={(a)}] 180:1.25) {};
		\node[vert] (j) at ([shift={(a)}] 270:1.25) {};
		\node[vert] (k) at ([shift={(i)}] 270:1.25) {};
		\node[vert,label={above right:$w_2$}] (l) at ([shift={(e)}] 0:1.25) {};

		\node[vert] (m) at ([shift={(e)}] 315:1.25) {};
		\node[vert] (n) at ([shift={(l)}] 325:1.25) {};

		\draw[sol,->, bend right=15] (a) to (b);
		\draw[sol,->, bend right=0] (b) to (d);
		\draw[sol,->, bend right=15] (d) to (e);
		\draw[sol,->, bend right=15] (e) to (f);
		\draw[sol,->, bend right=0] (f) to (g);
		\draw[sol,->, bend right=15] (g) to (a);
		\draw[sol,->] (i) to (a);
		\draw[sol,->] (l) to (e);

		\draw[snake] (j) to node[midway, label={left:\small{$4$}}] {} (a);
		\draw[snake] (k) to node[midway, label={left:\small{$5$}}] {} (i);
		\draw[snake] (m) to node[midway, label={right:\small{$4$}}] {} (e);
		\draw[snake] (n) to node[midway, label={right:\small{$5$}}] {} (l);

		\node[draw=none,inner sep=0] (bt) at ([shift={(b)}] 135:0.15) {};
		\node[draw=none,inner sep=0] (dt) at ([shift={(d)}] 135:0.15) {};
		\draw[snake] (bt) to (dt);

		\node[draw=none,inner sep=0] (ft) at ([shift={(f)}] 135:0.15) {};
		\node[draw=none,inner sep=0] (gt) at ([shift={(g)}] 135:0.15) {};
		\draw[snake] (gt) to (ft);

		\node[draw=none] (tmp) at (0,-2) {}; 

	\end{tikzpicture}
	\subcaption{Paths of $c_p$ before coloring. There are two paths: the first one from $w_1$ to $v_1$ and the second one from $w_2$ to $v_2$.}
\end{subfigure}
\quad
\begin{subfigure}[t]{0.47\textwidth}
	\centering
	\begin{tikzpicture}
		\node[vert] (a) at (0,0) {};
		\node[vert] (b) at ([shift={(a)}] 15:1.25) {};
		\node[vert] (d) at ([shift={(b)}] 45:1.25) {};
		\node[vert] (e) at ([shift={(d)}] 75:1.25) {};
		\node[vert] (f) at ([shift={(e)}] 195:1.25) {};
		\node[vert] (g) at ([shift={(f)}] 225:1.25) {};

		\node[vert] (i) at ([shift={(a)}] 180:1.25) {};
		\node[vert] (j) at ([shift={(a)}] 270:1.25) {};
		\node[vert] (k) at ([shift={(i)}] 270:1.25) {};
		\node[vert] (l) at ([shift={(e)}] 0:1.25) {};

		\node[vert] (m) at ([shift={(e)}] 315:1.25) {};
		\node[vert] (n) at ([shift={(l)}] 325:1.25) {};

		\draw[sol,->, bend right=15] (a) to node[midway, label={below:\small{$5$}}] {} (b);
		\draw[sol,->, bend right=0] (b) to node[midway, label={below right:\small{$45$}}] {} (d);
		\draw[sol,->, bend right=15] (d) to node[midway, label={left:\small{$5$}}] {} (e);
		\draw[sol,->, bend right=15] (e) to node[midway, label={above:\small{$5$}}] {} (f);
		\draw[sol,->, bend right=0] (f) to node[midway, label={above left:\small{$45$}}] {} (g);
		\draw[sol,->, bend right=15] (g) to node[midway, label={above left:\small{$5$}}] {} (a);
		\draw[sol,->] (i) to node[midway, label={above:\small{$4$}}] {} (a);
		\draw[sol,->] (l) to node[midway, label={above:\small{$4$}}] {} (e);

		\draw[snake] (j) to node[midway, label={left:\small{$4$}}] {} (a);
		\draw[snake] (k) to node[midway, label={left:\small{$5$}}] {} (i);
		\draw[snake] (m) to node[midway, label={right:\small{$4$}}] {} (e);
		\draw[snake] (n) to node[midway, label={right:\small{$5$}}] {} (l);

		\node[draw=none,inner sep=0] (bt) at ([shift={(b)}] 135:0.15) {};
		\node[draw=none,inner sep=0] (dt) at ([shift={(d)}] 135:0.15) {};
		\draw[snake] (bt) to (dt);

		\node[draw=none,inner sep=0] (ft) at ([shift={(f)}] 135:0.15) {};
		\node[draw=none,inner sep=0] (gt) at ([shift={(g)}] 135:0.15) {};
		\draw[snake] (gt) to (ft);

		\node[draw=none] (tmp) at (0,-2) {};

	\end{tikzpicture}
	\subcaption{Preliminary coloring of paths of $c_p$. There is a monochromatic cycle in color $5$}
\end{subfigure}
\\
\begin{subfigure}[t]{0.47\textwidth}
	\centering
	\begin{tikzpicture}
		\node[vert] (a) at (0,0) {};
		\node[vert] (b) at ([shift={(a)}] 15:1.25) {};
		\node[vert] (d) at ([shift={(b)}] 45:1.25) {};
		\node[vert] (e) at ([shift={(d)}] 75:1.25) {};
		\node[vert] (f) at ([shift={(e)}] 195:1.25) {};
		\node[vert] (g) at ([shift={(f)}] 225:1.25) {};

		\node[vert] (i) at ([shift={(a)}] 180:1.25) {};
		\node[vert] (j) at ([shift={(a)}] 270:1.25) {};
		\node[vert] (k) at ([shift={(i)}] 270:1.25) {};
		\node[vert] (l) at ([shift={(e)}] 0:1.25) {};

		\node[vert] (m) at ([shift={(e)}] 315:1.25) {};
		\node[vert] (n) at ([shift={(l)}] 325:1.25) {};

		\draw[sol,->, bend right=15] (a) to node[midway, label={below:\small{$5$}}] {} (b);
		\draw[sol,->, bend right=0] (b) to node[midway, label={below right:\small{$45$}}] {} (d);
		\draw[sol,->, bend right=15] (d) to node[midway, label={left:\small{$4$}}] {} (e);
		\draw[sol,->, bend right=15] (e) to node[midway, label={above:\small{$5$}}] {} (f);
		\draw[sol,->, bend right=0] (f) to node[midway, label={above left:\small{$45$}}] {} (g);
		\draw[sol,->, bend right=15] (g) to node[midway, label={above left:\small{$5$}}] {} (a);
		\draw[sol,->] (i) to node[midway, label={above:\small{$4$}}] {} (a);
		\draw[sol,->] (e) to node[midway, label={above:\small{$5$}}] {} (l);

		\draw[snake] (j) to node[midway, label={left:\small{$4$}}] {} (a);
		\draw[snake] (k) to node[midway, label={left:\small{$5$}}] {} (i);
		\draw[snake] (m) to node[midway, label={right:\small{$4$}}] {} (e);
		\draw[snake] (n) to node[midway, label={right:\small{$5$}}] {} (l);

		\node[draw=none,inner sep=0] (bt) at ([shift={(b)}] 135:0.15) {};
		\node[draw=none,inner sep=0] (dt) at ([shift={(d)}] 135:0.15) {};
		\draw[snake] (bt) to (dt);

		\node[draw=none,inner sep=0] (ft) at ([shift={(f)}] 135:0.15) {};
		\node[draw=none,inner sep=0] (gt) at ([shift={(g)}] 135:0.15) {};
		\draw[snake] (gt) to (ft);

	\end{tikzpicture}
	\subcaption{We recolor edge $(u_1,v_1)$ and path from $v_1$ to $w_2$. All edges are now safe}
\end{subfigure}
  \caption{Example of algorithm described in Lemma \ref{lemma:coloring_cycle_of_paths}}
  \label{fig:path_recoloring}
\end{figure}
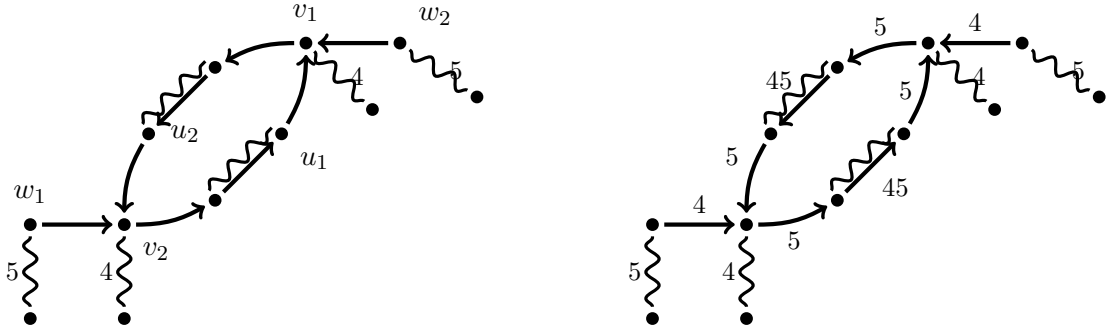
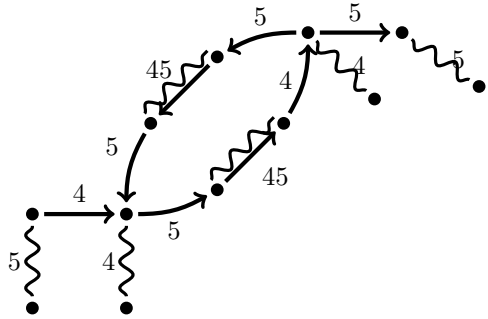

\begin{lemma}
\label{lemma:coloring_cycle_of_paths}
Let $c_p$ be a directed cycle of $G_p$ considered at some step of Algorithm Color $G_2$. Then it is possible to color each edge belonging to the surrounding of each path of $G'_2$ occurring on $c_p$ in such a way that it is safe.
\end{lemma}
\dowod
Suppose that the cycle $c_p$ goes through vertices $p_1, p_2, \ldots, p_k$ in this order. Let $(u_i, v_i)$ denote the border of path $p_i$ of $G'_2$ for each $i, 1 \leq i \leq k$.  We start by coloring the path $p_1$ and its surrounding in the manner described in the proof of Lemma \ref{path45}. If the border $(u_1, v_1)$ is not double, then we leave it uncolored. Next we color each of the paths $p_2, \ldots, p_{k-1}$ and their surroundings together with their borders, also in the way described in the proof of Lemma \ref{path45}. Next we have to check two possibilities of dealing with the path $p_k$. First we  color the path $p_k$ together with  its border and surrounding in the same manner
as the remaining paths $p_1, \ldots, p_{k-1}$ and if the border $(u_1, v_1)$ is uncolored, because it is not double, we color it with the only possible color of $\Kd$. It may happen, however, that by doing so we create a monochromatic cycle $c'$ that is formed by the part of $p_1$ between 
$v_1$ and $v_2$, the part of $p_2$ between $v_2$ and $v_3$ and so on until the part of $p_k$ between $v_k$ and $v_1$. If this is the case, then we leave the part of $p_k$ between $v_k$ and $v_1$ colored as it is and uncolor the remaining part of  $p_k$. If the border $(u_1, v_1)$
is not double, then we change its color to the opposite one. If the border $(u_1, v_1)$ is double, then we change the color of the edge proceeding it on $p_1$ to the opposite one. Next we change the orientation of the uncolored part of $p_k$ as follows. The endpoints of $p_k$ are $v_k$ and some vertex $w_k$ and originally $p_k$ is oriented from $w_k$ to $v_k$. Now we change the orientation of the part $p'$ of $p_k$ between $v_1$ and $w_k$ so that it is directed from $v_1$ to $w_k$.  Let $e=(w, v_1)$ denote the edge of $p'$ incident to $v_1$.  Since $v_1$ has degree $4$ in the multigraph $G'_2$, there exists only one color $k$ of $\Kd$ that can be used for coloring $e$. The rest of $p'$ is colored in the standard way. We only have to show that the edge $e$ is safe, as every other edge considered in this lemma is safe by reasoning analogous to that used in two previous lemmas. The edge $e$ is safe because it is colored with same color $k$ that every edge of $c'$ but one is colored with. Also, $e$ is the only edge incident to $c'$ but not lying on $c'$ that is colored with $k$. The example of this algorithm is presented on Figure \ref{fig:path_recoloring}.

\koniec

\section{The proof of Lemma \ref{F12}} \label{dowodf12}
First we want to guarantee that property 6. is satisfied. Let us say that an edge $e$ is a {\bf \em d-edge} if it belongs to $M$ and some kite. Let $S$ contain every cycle of $C_2$ of odd length $l$ that has $l$ different incident d-edges. Let $T$ denote the set of all d-edges.
We build a bipartite graph $G_d=(S \cup T, E_d)$ such that there exists an edge in $E_d$ between a cycle $c$ of $S$ and  edge $e$ of $T$ iff $e$ is incident to $c$. Furthermore for each 4-kite $s$ incident to at most three cycles in $S$ we merge vertices corresponding to d-edges in s into one vertex. Let us notice that the degree of each d-edge of $T$ in $G_d$ is at most $3$ and the degree of each cycle $c$ of $S$ is at least $3$. We compute a matching $M_d$ of size $|S|$ in the graph $G_d$. By Hall's Theorem such a matching always exists. Then for each cycle $c$ and matched to it d-edge $e$ we will either (i) add an outgoing edge of $c$ incident to $e$ to $F_2$ or (ii) ensure that $e$ is not a double edge. 

We begin with the proof for the case when there are only $3$-kites. 

Let $t$ be any $3$-kite on vertices $u,v,w$ such that $e=(u,v)$ is a d-edge of $t$ and $e_2=(u,w)$ and $e_3=(v,w)$.

We begin with the case when $t$ has three incoming and three outgoing edges of $C_2$ incident to it. We add $e$ to $F_1$. To $F_2$ we add any outgoing edge of $t$ belonging to $C_2$ (it can be an edge required by $M_d$).  If at some later point $t$ gets three incident incoming edges of $F_2$, then we remove $e$ from $F_1$ and replace it with that one of the edges $e_2, e_3$ that is incident to an outgoing edge of $F_2$ and orient it so that it is directed to $w$. We do it so that Lemma \ref{wlasnosci} Property (2) is satisfied.

Whenever $e$ does not belong to $Z$, we add it to $F_1$ and an outgoing edge of $t$ incident to $e$ to $F_2$.

\begin{figure}[h!]
	\centering
\begin{subfigure}[t]{0.47\textwidth}
	\centering
	\begin{tikzpicture}
		\trikite{}{}
		\trExta{}{}
		\tlExta{$f$}{}
		\tInternal
		\lLabel{$e_2$}
		\rLabel{$e_1$}
		\blExta{}{}
		\blExtb{}{}
		\tLabel{$e$}
	\end{tikzpicture}
	\subcaption{We add $f$ to $F_2$ and $e_2$ to $F_1$. In $G_2'$ we orient $e_2$ so that it is directed from a common endpoint with $e$.}
\end{subfigure}
\quad
\begin{subfigure}[t]{.47\textwidth}
	\centering
	\begin{tikzpicture}
		\trikite{}{}
		\trExta{$f_1$}{}
		\trExtb{}{}
		\tlExta{$f_2$}{}
		\lLabel{$e_2$}
		\rLabel{$e_1$}
		\blExtb{$f_3$}{}
		\ltHalfEdge
		\rbHalfEdge
		\tLabel{$e$}
	\end{tikzpicture}
	\subcaption{We add $f_1$ to $F_2$ and $e$ to $F_1$. If $e_2$ is in $Z$ we orient it from $w$ to $u$ and make $f_2$ outgoing. If $e_3$ is in $Z$ we orient it from $v$ to $w$ and make $f_3$ outgoing}
\end{subfigure}
\\
\begin{subfigure}[t]{0.47\textwidth}
	\centering
	\begin{tikzpicture}
		\trikite{}{}
		\trExta{}{}
		\tlExta{$f_1$}{}
		\lLabel{$e_2$}
		\rLabel{$e_1$}
		\blExta{}{}
		\blExtb{}{}
		\trHalfEdge
		\ltHalfEdge
		\tLabel{$e$}
	\end{tikzpicture}
	\subcaption{If $e_2$ is in $Z$ we make $f_1$ outgoing. In this case we add $e$ to $F_1$ and $f_1$ to $F_2$ and orient $e_2$ from $u$ to $w$. If $e$ is in $Z$ we make $f_2$ outgoing, add $e_3$ to $F_1$ and $f_2$ to $F_2$ and we orient $e_3$ from $v$ to $w$}
\end{subfigure}
\quad
\begin{subfigure}[t]{0.47\textwidth}
	\centering
	\begin{tikzpicture}
		\trikite{}{}
		\trExta{$f_1$}{}
		\trExtb{}{}
		\blExta{$f_2$}{}
		\lLabel{$e_2$}
		\rLabel{$e_1$}
		\blExtb{}{}
		\tlHalfEdge
		\ltHalfEdge
		\tLabel{$e$}
	\end{tikzpicture}
	\subcaption{If $e$ is in $Z$ we add $e_2$ to $F_1$ and $f_1$ to $F_2$. If $e_2$ is in $Z$ we add $e$ to $F_1$ and $f_2$ to $F_2$}
\end{subfigure}
\\
\begin{subfigure}[t]{0.47\textwidth}
	\centering
	\begin{tikzpicture}
		\trikite{}{}
		\trExta{$f_3$}{}
		\tlExta{$f_2$}{}
		\lLabel{$e_2$}
		\rLabel{$e_1$}
		\tlExtb{}{}
		\blExtb{$f_1$}{}
		\trHalfEdge
		\lbHalfEdge
		\tLabel{$e$}
	\end{tikzpicture}
	\subcaption{If $e_2$ is in $Z$ we make $f_1$ outgoing, add $e$ to $F_1$ and $f_2$ to $F_2$. If $e$ is in $Z$ we make $f_3$ outgoing, add $e_3$ to $F_1$ and $f_2$ to $F_2$}
\end{subfigure}
\quad
\begin{subfigure}[t]{0.47\textwidth}
	\centering
	\begin{tikzpicture}
		\trikite{}{}
		\trExta{}{}
		\tlExta{$f_1$}{}
		\lLabel{$e_2$}
		\rLabel{$e_1$}
		\tlExtb{}{}
		\blExtb{}{}
		\lInternal
		\tLabel{$e$}
	\end{tikzpicture}
	\subcaption{We add $e$ to $F_1$ and $f_1$ to $F_2$}
\end{subfigure}

	\caption{Assigning edges of $3$-kites with two incoming and two outgoing edges}
	\label{fig:triangles_2out2in}
\end{figure}
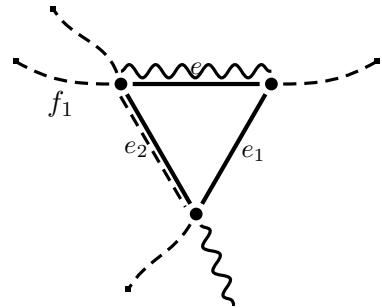

 We consider now the cases when $t$ has two incoming and two outgoing edges
of $C_2$ incident to it. In  Figure \ref{fig:triangles_2out2in} we show how to assign edges of all possible $3$-kites with two incoming and two outgoing edges to $F_1$ and $F_2$.

In case of triangles with one incoming and one outgoing edge there are already two edges of $t$ in $C_2 \cup Z$. Therefore we add remaining edge to $F_1$ and incident edge (we can guarantee that it is outgoing when constructing $D_2$) to $F_2$.

To finish the proof we must consider all cases for $4$-kites.
We say that an edge $e$ is a {\bf \em side edge} if it belongs to 4-kite, but not to $M$. Let $c$ be any $4$-kite on vertices $u,v,w,z$ such that $(u,v)$ and $(w,z)$ are d-edges and $(u,z)$ and $(v,w)$ are side edges. Let $I_c$ be the set of those edges in $I(C_2)$, whose both endpoints are in $\{u,v,w,z\}$ (so $I(C_2) \setminus I_c$ contains all incoming and outgoing edges incident to $c$). First we consider the cases when $c$ has one incoming and one outgoing edge of $C_2$ incident to it (by our construction of $C_2$ these edges must be incident to different vertices of $c$). If these edges are not incident to the same side edge, we add outgoing edge to $F_2$ and the side edge incident to it to $F_1$. Otherwise let's assume that they are incident to $(u,z)$, and that edge incident to $u$ is outgoing in $D_2$, whereas edge incident to $z$ is outgoing in $opp(D_2)$.
\begin{enumerate}
	\item $u$ and $z$ are incident to at most $2$ half-edges or $1$ edge of $I_c$ and $1$ half-edge - we divide half-edges into $Z_1$ and $Z_2$ so that degree of $z$ in $I(C_2) \cup Z_1 \cup M$ is 3 and degree of $u$ in $I(C_2) \cup Z_2 \cup M$ is 3. Then we add $(u,z)$ to $F_1$ and outgoing edge to $F_2$.
	\item $z$ is incident to $1$ edge of $I_c$ and $2$ half-edges and $(u,z)$ is not in $I_c$ - then half-edge $(u,z)$ is in $C_2$ so $u$ is not incident to any edge in $I_c$. We divide half-edges into $Z_1$ and $Z_2$ so that $(u,v)$ is not in $Z_1$ and degree of $u$ in $I(C_2) \cup Z_2 \cup M$ is 3. If outgoing edge is incident to $u$ we add $(u,v)$ to $F_1$, and otherwise we add $(u,z)$ to $F_1$. In both cases we add outgoing edge to $F_2$.
	\item $(u,z)$ is in $I_c$ - we divide half-edges so that $(u,v)$ is not in $Z_1$ and $(w,z)$ is not in $Z_2$. We add d-edge incident to outgoing edge to $F_1$ and outgoing edge to $F_2$.
\end{enumerate}

Now let's consider the cases when two vertices incident to the same d-edge, say $(u,v)$, are incident to one incoming and one outgoing edge each and the other two vertices aren't incident to any incoming or outgoing edges. If $w$ and $z$ are incident to two half-edges each then two half-edges incident to $w$ go to $Z_1$, and the other two half-edges go to $Z_2$. If $w$ and $z$ are incident to one half-edge each then $I(C_2)$ must contain $(w,z)$ and we divide half-edges into $Z_1$ and $Z_2$ arbitrarily. There are three cases depending on which edges incident to $z$ are in $Z \cup I(C_2)$ (cases when in $Z$ there are half-edges incident to $w$ are symmetric):
\begin{enumerate}
	\item $(w,z)$ and $(v,z)$ are in $Z \cup I(C_2)$ - we add outgoing edge incident to $u$ and $(v,z)$ to $F_2$ and add $(u,z)$ and $(v,w)$ to $F_1$.
	\item $(w,z)$ and $(u,z)$ are in $Z \cup I(C_2)$ - we add outgoing edge incident to $v$ to $F_2$ and add $(v,w)$ to $F_1$.
	\item $(v,z)$ and $(u,z)$ are in $Z \cup I(C_2)$ - we add outgoing edge incident to $v$ to $F_2$. If incoming edge incident to $v$ is also in $F_2$ we add $(v,w)$ to $F_1$. Otherwise we add $(w,z)$ to $F_1$
\end{enumerate}

The next cases are similar to the previous one, but now vertices incident to one incoming and one outgoing edge each are incident to the same side edge, say $(u,z)$. We divide half-edges same as before, so two half-edges incident to $v$ go to $Z_1$, and the other two go to $Z_2$. Now the cases are:
\begin{enumerate}
	\item $(v,z)$ and $(v,w)$ are in $Z \cup I(C_2)$ and $z$ was matched in $M_d$ with $(w,z)$ - we add outgoing edge incident to $z$ and $(v,z)$ to $F_2$. If incoming edge incident to $z$ is in $F_2$ we add $(w,z)$ and $(u,z)$ to $F_1$. Otherwise we add $(w,z)$ and $(v,w)$ to $F_1$.
	\item $(v,z)$ and $(v,w)$ are in $Z \cup I(C_2)$ and $z$ wasn't matched in $M_d$ with $(w,z)$ - we add outgoing edge incident to $u$ and $(v,z)$ to $F_2$ and we add $(u,v)$ and $(z,w)$ to $F_1$.
	\item $(v,z)$ and $(u,v)$ are in $Z \cup I(C_2)$ - we add outgoing edge incident to $u$ and $(v,z)$ to $F_2$. If incoming edge incident to $u$ is in $F_2$ we add $(u,z)$ and $(v,w)$ to $F_1$. Otherwise we add $(v,w)$ and $(w,z)$ to $F_1$.
	\item $(u,v)$ and $(v,w)$ are in $Z \cup I(C_2)$ and $z$ was matched in $M_d$ with $(w,z)$ - we add outgoing edge incident to $z$ to $F_2$ and $(w,z)$ to $F_1$.
	\item $(u,v)$ and $(v,w)$ are in $Z \cup I(C_2)$ and $z$ wasn't matched in $M_d$ with $(w,z)$ - we add outgoing edge incident to $u$ to $F_2$. If incoming edge incident to $u$ is in $F_2$ we add $(u,z)$ to $F_1$. Otherwise we add $(w,z)$ to $F_1$.
\end{enumerate}

Now there are three cases in which there is a vertex in $c$, say $u$, incident to two edges in $I(C_2)\setminus I_c$, two vertices incident to one edge in $I(C_2) \setminus I_c$ each and a vertex incident to no edge in $I(C_2) \setminus I_c$. The first case is when $z$ isn't incident to any edge in $I(C_2) \setminus I_c$. Let's assume that edge incident to $v$ is outgoing in $D_2$. We divide half-edges into $Z_1$ and $Z_2$ so that (i) in $Z_1 \cup I(C_2)$ there is an edge incident to $v$ and edge incident to $z$ and, similarly, in $Z_2 \cup I(C_2)$ there is an edge incident to $w$ and edge incident to $z$ (ii) no two half-edges incident to $u$ are in the same set (this condition can be satisfied because in $C_2$ there are at most two half-edges incident to $u$). Now we consider all subcases of which edges are in $Z \cup I(C_2)$:
\begin{enumerate}
	\item $(v,w)$ and $(w,z)$ are in $Z \cup I(C_2)$ - we add outgoing edge incident to $u$ to $F_2$ and $(u,z)$ to $F_1$.
	\item $(v,w)$ and $(v,z)$ are in $Z \cup I(C_2)$ - we add outgoing edge incident to $u$ and $(v,z)$ to $F_2$ and add $(u,z)$ and $(w,z)$ to $F_1$.
	\item $(v,w)$ and $(u,z)$ are in $Z \cup I(C_2)$ - we add outgoing edge incident to $u$ to $F_2$. If incoming edge incident to $u$ is also in $F_2$ we add $(u,v)$ to $F_1$. Otherwise we add $(w,z)$ to $F_1$.
	\item $(v,z)$ and $(w,z)$ are in $Z \cup I(C_2)$ - we add outgoing edge incident to $u$ to $F_2$ and $(u,v)$ to $F_1$.
	\item $(v,z)$ and $(u,z)$ are in $Z \cup I(C_2)$ - there is no edge incident to $w$, so edge incident to $v$ is outgoing. We add outgoing edge incident to $v$ to $F_2$ and $(v,w)$ to $F_1$.
	\item $(u,v)$ and $(w,z)$ are in $Z \cup I(C_2)$ - we add outgoing edge incident to $u$ to $F_2$ and $(u,z)$ to $F_1$.
	\item $(u,v)$ and $(v,z)$ are in $Z \cup I(C_2)$ - we add outgoing edge incident to $u$ and $(v,z)$ to $F_2$ and add $(u,z)$ and $(w,z)$ to $F_1$.
	\item $(u,w)$ and $(w,z)$ are in $Z \cup I(C_2)$ - we add outgoing edge incident to $u$ and $(u,w)$ to $F_2$ and add $(u,v)$ and $(u,z)$ to $F_1$.
	\item $(u,w)$ and $(v,w)$ are in $Z \cup I(C_2)$ - $(u,w)$ is incident to neither $v$ nor $z$, so edge incident to $w$ is outgoing. We add outgoing edge incident to $u$ to $F_2$. If incoming edge incident to $u$ is also in $F_2$ we add $(u,z)$ to $F_1$. Otherwise we add $(w,z)$ to $F_1$.
	\item $(w,z)$ and $(u,z)$ are in $Z \cup I(C_2)$ - we add outgoing edge incident to $u$ to $F_2$ and $(u,v)$ to $F_1$.
\end{enumerate}

In the second case $w$ isn't incident to any edge in $I(C_2) \setminus I_c$. We divide half-edges same as in the previous case, so we guarantee that (i) in $Z_1 \cup I(C_2)$ there is an edge incident to $v$ and edge incident to $w$ and, in $Z_2 \cup I(C_2)$ there is an edge incident to $z$ and edge incident to $w$. Condition (ii) remains the same. Now the subcases are as follows:
\begin{enumerate}
	\item $(v,w)$ and $(w,z)$ are in $Z \cup I(C_2)$ - we add outgoing edge incident to $u$ to $F_2$ and $(u,v)$ to $F_1$.
	\item $(v,w)$ and $(u,w)$ are in $Z \cup I(C_2)$ - there is no edge incident to $z$, so edge incident to $v$ is outgoing. We add outgoing edge incident to $u$ to $F_2$. If incoming edge incident to $u$ is also in $F_2$ we add $(u,z)$ to $F_1$. Otherwise we add $(w,z)$ to $F_1$.
	\item $(v,z)$ and $(v,w)$ are in $Z \cup I(C_2)$ - we add outgoing edge incident to $u$ and $(v,z)$ to $F_2$ and add $(u,z)$ and $(w,z)$ to $F_2$.
	\item $(v,z)$ and $(u,w)$ are in $Z \cup I(C_2)$ - we add outgoing edge incident to $u$ to $F_2$. If incoming edge incident to $u$ is also in $F_2$ we add $(u,v)$ to $F_1$, and otherwise we add $(w,z)$ to $F_1$. In this case at least one of $(u,v), (w,z)$ is not double edge, so cycle going through $v$ and $z$ satisfies condition 6 from lemma.
	\item $(v,z)$ and $(w,z)$ are in $Z \cup I(C_2)$ - we add outgoing edge incident to $u$ and $(v,z)$ to $F_2$ and add $(u,z)$ and $(v,w)$ to $F_2$.
	\item $(u,v)$ and $(v,w)$ are in $Z \cup I(C_2)$ - there is no edge incident to $z$, so edge incident to $v$ is outgoing. We add outgoing edge incident to $u$ to $F_2$. If incoming edge incident to $u$ is also in $F_2$ we add $(u,z)$ to $F_1$. Otherwise we add $(w,z)$ to $F_1$.
	\item $(u,v)$ and $(w,z)$ are in $Z \cup I(C_2)$ - we add outgoing edge incident to $u$ to $F_2$. If incoming edge incident to $u$ is also in $F_2$ we add $(u,z)$ to $F_1$. Otherwise we add $(v,w)$ to $F_1$.
	\item $(w,z)$ and $(u,w)$ are in $Z \cup I(C_2)$ - there is no edge incident to $v$, so edge incident to $z$ is outgoing. We add outgoing edge incident to $u$ to $F_2$ and $(u,v)$ to $F_1$.
	\item $(u,z)$ and $(v,w)$ are in $Z \cup I(C_2)$ - $(u,z)$ is incident to neither $v$ nor $w$, so edge incident to $z$ is outgoing. We add outgoing edge incident to $u$ to $F_2$. If incoming edge incident to $u$ is also in $F_2$ we add $(u,v)$ to $F_1$. Otherwise we add $(w,z)$ to $F_1$.
	\item $(u,z)$ and $(w,z)$ are in $Z \cup I(C_2)$ - there is no edge incident to $v$, so edge incident to $z$ is outgoing. We add outgoing edge incident to $u$ to $F_2$. If incoming edge incident to $u$ is also in $F_2$ we add $(u,v)$ to $F_1$. Otherwise we add $(v,w)$ to $F_1$.
\end{enumerate}

In the third case $v$ isn't incident to any edge in $I(C_2) \setminus I_c$. Similarly as before we divide half-edges to guarantee that (i) in $Z_1 \cup I(C_2)$ there is an edge incident to $w$ and edge incident to $v$ and, in $Z_2 \cup I(C_2)$ there is an edge incident to $z$ and edge incident to $v$. Once again condition (ii) remains the same. Now the subcases are as follows:
\begin{enumerate}
	\item $(v,w)$ and $(v,z)$ are in $Z \cup I(C_2)$ - we add outgoing edge incident to $u$ to $F_2$ and $(u,v)$ to $F_1$. Condition 6 from lemma is satisfied, because $(w,z)$ is not a double edge.
	\item $(v,w)$ and $(u,v)$ are in $Z \cup I(C_2)$ - we add outgoing edge incident to $u$ to $F_2$. If incoming edge incident to $u$ is also in $F_2$ we add $(u,z)$ to $F_1$. Otherwise we add $(w,z)$ to $F_1$.
	\item $(u,w)$ and $(v,w)$ are in $Z \cup I(C_2)$ - we add outgoing edge incident to $u$ and $(v,w)$ to $F_2$. If incoming edge incident to $u$ is also in $F_2$ we add $(u,z)$ and $(u,v)$ to $F_1$. Otherwise we add $(u,v)$ and $(w,z)$ to $F_1$.
	\item $(u,w)$ and $(v,z)$ are in $Z \cup I(C_2)$ - $(u,w)$ is incident to neither $v$ nor $z$, so edge incident to $w$ is outgoing. We add $u$ to $F_2$ and $(u,v)$ to $F_1$.
	\item $(w,z)$ and either $(v,w)$ or $(v,z)$ are in $Z \cup I(C_2)$ - we add outgoing edge incident to $u$ to $F_2$ and $(u,v)$ to $F_1$.
	\item $(w,z)$ and $(u,v)$ are in $Z \cup I(C_2)$ - we add outgoing edge incident to $u$ to $F_2$. If incoming edge incident to $u$ is also in $F_2$ we add $(u,z)$ to $F_1$. Otherwise we add $(v,w)$ to $F_1$.
	\item $(v,z)$ and $(u,v)$ are in $Z \cup I(C_2)$ - we add outgoing edge incident to $u$ and $(v,z)$ to $F_2$. If incoming edge incident to $u$ is also in $F_2$ we add $(u,z)$ and $(w,z)$ to $F_1$. Otherwise we add $(v,w)$ and $(w,z)$ to $F_1$.
	\item $(u,z)$ and $(v,w)$ are in $Z \cup I(C_2)$ - $(u,z)$ is incident to neither $v$ nor $w$, so edge incident to $z$ is outgoing. We add outgoing edge incident to $u$ to $F_2$ and $(u,v)$ to $F_1$.
	\item $(u,z)$ and $(v,z)$ are in $Z \cup I(C_2)$ - $(u,z)$ is incident to neither $v$ nor $w$, so edge incident to $z$ is outgoing. We add outgoing edge incident to $u$ and $(v,z)$ to $F_2$ and add $(u,v)$ and $(v,w)$ to $F_1$.
\end{enumerate}

The final case when $c$ is incident to two incoming and two outgoing edges of $C_2$ is when each vertex of $c$ is incident to one edge of $I(C_2) \ I_c$. First suppose that $(u,z)$ is in $I(C_2)$. Then we can assume that in $Z$ there is half-edge incident to $v$ (or there is $(v,w)$ in $I(C_2)$), because other cases are symmetric:
\begin{enumerate}
	\item $(u,v)$ is in $Z \cup I(C_2)$ - we add outgoing edge incident to $u$ or $z$ to $F_2$ and $(w,z)$ to $F_1$.
	\item $(v,z)$ is in $Z \cup I(C_2)$ - we add outgoing edge incident to $u$ or $z$ and $(v,z)$ to $F_2$ and $(u,v)$ and $(w,z)$ to $F_1$.
	\item $(v,w)$ is in $Z \cup I(C_2)$ - we add outgoing edge incident to $u$ or $z$ to $F_2$ and d-edge adjacent to added outgoing edge to $F_1$.
\end{enumerate}
If $(u,w)$ is in $I(C_2)$ we assume that in $Z \cup I(C_2)$ there is an edge incident to $z$ and that edge incident to $z$ is outgoing. If edge incident to $u$ is outgoing we add it to $F_2$ and add $(u,v)$ to $F_1$. If edge incident to $w$ is outgoing then there are three cases:
\begin{enumerate}
	\item $(u,z)$ is in $Z \cup I(C_2)$ - we add outgoing edge incident to $w$ and $(u,w)$ to $F_2$ and add $(v,w)$ and $(w,z)$ to $F_1$.
	\item $(v,z)$ is in $Z \cup I(C_2)$ - we add outgoing edge incident to $w$ to $F_2$ and add $(w,z)$ to $F_1$.
	\item $(w,z)$ is in $Z \cup I(C_2)$ - we add outgoing edge incident to $w$ to $F_2$ and add $(u,v)$ to $F_1$.
\end{enumerate} 
Now suppose that $(u,v)$ is in $I(C_2)$. We assume that in $Z \cup I(C_2)$ there is an edge incident to $z$ and that edge incident to $z$ is outgoing:
\begin{enumerate}
	\item $(u,z)$ or $(v,z)$ is in $Z \cup I(C_2)$ - we add outgoing edge incident to $z$ to $F_2$ and $(w,z)$ to $F_1$.
	\item $(w,z)$ is in $Z \cup I(C_2)$ - we add outgoing edge incident $z$ to $F_2$ and $(u,z)$ to $F_1$.
\end{enumerate}

Finally suppose that there are no whole edges inside c, so there are four half-edges. Into $Z_1$ belong half-edges adjacent to outgoing edges in $D_2$ and into $Z_2$ those adjacent to outgoing edges in $opp(D_2)$. Now we have to consider all possible edges in $Z$:
\begin{enumerate}
	\item $(u,z)$ and one other edge is in $Z$ - we act the same as in case with side edge.
	\item $(u,v)$ and $(v,z)$ are in $Z$ - if edge incident to $z$ is outgoing we add it to $F_2$. Otherwise edge incident to $v$ is outgoing and we add it to $F_2$. In both cases we add $(w,z)$ to $F_1$.
	\item $(u,v)$ and $(w,z)$ are in $Z$ - we add any outgoing edge to $F_2$ and adjacent side edge to $F_1$.
	\item $(u,w)$ and $(v,z)$ are in $Z$ - we add any outgoing edge to $F_2$ and adjacent d-edge to $F_1$.
\end{enumerate}

Now let's consider the case when $c$ has three incoming and three outgoing edges of $C_2$ incident to it and there is a vertex, say $z$ which is not incident to any incoming or outgoing edge. Then $z$ is incident to two half-edges, on of which is in $Z$. If in $Z$ there is half-edge $(w,z)$, we add outgoing edge incident to $w$ to $F_2$ and add $(u,z)$ to $F_1$. If incoming edge incident to $w$ is also in $F_2$ we add $(v,w)$ to $F_1$, so that property 5 from lemma is satisfied. If in $Z$ there is either $(v,z)$ or $(u,z)$, we add outgoing edge incident to $w$ to $F_2$ and $(w,z)$ to $F_1$.

In all other cases when $c$ has three incoming and three outgoing edges of $C_2$ incident to it, there are at most two half-edges, each incident to different vertex. In these cases we divide half-edges into $Z_1$ and $Z_2$ in such way, that to $Z_1$ belongs half-edge incident to outgoing edge in $D_2$. Now we have to consider all cases to which vertices incoming and outgoing vertices are incident:
\begin{enumerate}
	\item Vertices incident to two edges of $I(C_2) \setminus I_c$ are incident to the same d-edge, say $(u,v)$ - let's assume that edge incident to $w$ is outgoing (and therefore either half-edge incident to $w$ is in $Z$ or $(w,z)$ is in $I(C_2)$). Then we add outgoing edge incident to $w$ to $F_2$ and either add $(v,w)$ to $F_1$ if $(w,z)$ is in $I(C_2) \cup Z$ or add $(w,z)$ to $F_1$ otherwise.
	\item Vertices incident to two edges of $I(C_2) \setminus I_c$ are incident to the same side edge say $(u,z)$ - let's assume that edge incident to $w$ is outgoing (and therefore either half-edge incident to $w$ is in $Z$ or $(v,w)$ is in $I(C_2)$). Then if $(v,w)$ is in $Z \cup I(C_2)$ we add outgoing edge incident to $w$ to $F_2$ and $(w,z)$ to $F_1$. If $(u,w)$ is in $Z$ we add outgoing edge incident to $u$ to $F_2$ and $(u,v)$ to $F_1$. Finally if $(w,z)$ is in $Z$ we add outgoing edge incident to $z$ to $F_2$ and $(v,w)$ to $F_1$. If incoming edge incident to $z$ is also in $F_2$ we add $(u,z)$ to $F_1$, so that property 5 from lemma is satisfied
	\item None of the above cases - let's assume that $u$ and $w$ are incident to two edges in $I(C_2) \setminus I_c$ and edge incident to $z$ is outgoing (and therefore either half-edge incident to $z$ is in $Z$ or $(v,z)$ is in $I(C_2)$). Then if $(v,z)$ is in $Z \cup I(C_2)$ we add outgoing edge incident to $z$ to $F_2$ and $(w,z)$ to $F_1$. If $(u,z)$ is in $I(C_2)$ we add outgoing edge incident to $u$ to $F_2$ and $(u,v)$ to $F_1$. Finally if $(w,z)$ is in $Z$ there are two subcases:
		\begin{enumerate}
			\item In matching $M_d$ cycle incident to vertex $u$ is matched to d-edge $(u,v)$ - then we add outgoing edge incident to $u$ to $F_2$ and $(u,v)$ to $F_1$
			\item Otherwise we add outgoing edge incident to $w$ to $F_2$. If also incoming edge incident to $w$ is in $F_2$ we add $(v,w)$ to $F_1$, and if it isn't in $F_2$ we add $(u,v)$ to $F_1$.
		\end{enumerate}
\end{enumerate}
 
In the case when $c$ is incident to four incoming and four outgoing edges of $C_2$ we add outgoing edge incident to $u$ or $v$ to $F_2$ (depending on which one of cycles incident to these vertices was matched to $(u,v)$ in $M_d$; if none of them we choose arbitrarily) and add $(u,v)$ to $F_1$.

\koniec

\section{Proof of Lemma \ref{lem:cycle_cover_optimality}}
\label{sec:c2_opt_proof}

Let's now see, that the cycle cover we have found using our gadgets is indeed
  what had been promised --- the maximum weight cycle cover (in which we agree
  to having paths ending with half-edges) not containing kites from $G_1$. To
  prove that we will show, that no such cycle cover of $G$ has been blocked by
  our gadgets and demands, so for every proper cycle cover of $G$, it can be
  translated into a b-matching in the modified graph. Let's start off with
  triangles.

  \begin{lemma}
    \label{lem:maxtsp_g2compute_trikite}
    Let $K \in \mathcal{C}_1$ be a 3-kite in the graph $G_1$. Let $C$ be a cycle
    cover of $G$ not containing $K$ (as one of the cycles). There exists a selection
    of edges in the gadget $\mathcal{G}_K$ corresponding to $K$, that is
    compliant with the cycle cover $\mathcal{C}$ and every node $v$ in
    $\mathcal{G}_K$ has exactly $b(v)$ adjacent edges in it. Its total weight
    will be equal to the weight of $\mathcal{C}$.
  \end{lemma}
  \dowod
    Since the cycle cover $C$ doesn't contain $K$ as one of its cycles, it will
    have at least two edges connecting the nodes of this triangle with other
    vertices in the graph (that are \emph{external} with regard to $K$). These
    edges are replicated in the gadget-modified graph, so there is no doubt,
    they can be selected into the b-matching. We will now present, how to handle
    the edges of the triangle $K$ and the gadget $\mathcal{G}_K$. We will
    consider different interactions between $\mathcal{C}$ and $K$.
\begin{itemize}
    \item If $K \cap \mathcal{C} = \emptyset$ (no edge of the kite is used
    in the cycle cover), then the demands of vertices $u$, $v$ and $w$ are
    fulfilled by the external edges. Additionally, we select the edges $\langle
    x_1, x_2\rangle$ (middle edge on the right side of the gadget), $\langle
    x_5, x_6\rangle$ (middle on the left side), $\langle x_3, p\rangle$ and
    $\langle x_4,q\rangle$.

    \item  If $K \cap \mathcal{C} = \{\<u,v\>\}$ (the cycle cover contains
    one side of the triangle), the b-matching obviously contains $\<u,x_1\>$ and
    $\<x_2,v\>$ --- the halves of the edge $\<u,v\>$. We also select the middle
    edges of two other sides of the triangle, namely $\<x_3,x_4\>$ and
    $\<x_5,x_6\>$. We satisfy the demands of $p$ and $q$ by connecting them with
    $x_1$ and $x_2$ respectively.

    \item  Finally, if $K \cap \mathcal{C} = \{\<u,v\>,\<v,w\>\}$ (the cycle
    cover contains two sides of the triangle), we select the corresponding
    half-edges $\<u,x_1\>$, $\<x_2,v\>$, $\<v,x_3\>$ and $\<x_4,w\>$. The nodes
    $p$ and $q$ are connected with $x_6$ and $x_5$.
		
\end{itemize}		
  \koniec

\begin{figure}[h]
    \centering

		\begin{center}
\begin{subfigure}{.32\textwidth}
  \begin{tikzpicture}
    \tikzset{every node/.style={draw,fill=black,circle,scale=.5,inner sep=0pt,outer sep=2pt}}
    \tikzstyle{pom}=[draw=black!60,thin]
    \node[vert]   (a) at (-1, -1) {};
    \node[vert]   (b) at (1, -1) {};
    \node[vert]   (c) at (1, 1) {};
    \node[vert]   (d) at (-1, 1) {};

    \node (aab) at (-.33,-1) {};
    \node (abb) at (.33, -1) {};

    \node (bbc) at (1, -.33) {};
    \node (bcc) at (1, .33) {};

    \node (ccd) at (.33, 1) {};
    \node (cdd) at (-.33, 1) {};

    \node (add) at (-1, .33) {};
    \node (aad) at (-1, -.33) {};

    \node (aac) at (-.7, -.7) {};
    \node (acc) at (.7, .7) {};
    \node (bbd) at (.7, -.7) {};
    \node (bdd) at (-.7, .7) {};

    \draw[pom] (a) -- (aab);
    \draw[pom] (abb) -- (b);
    \draw[pom] (b) -- (bbc);
    \draw[pom] (bcc) -- (c);
    \draw[pom] (c) -- (ccd);
    \draw[pom] (cdd) -- (d);
    \draw[pom] (d) -- (add);
    \draw[pom] (aad) -- (a);

    \draw[pom] (aab) -- (abb);
    \draw[sol] (bbc) -- (bcc);
    \draw[sol] (ccd) -- (cdd);
    \draw[sol] (add) -- (aad);

    \draw[pom] (a) -- (aac);
    \draw[pom] (b) -- (bbd);
    \draw[pom] (c) -- (acc);
    \draw[pom] (d) -- (bdd);

    \node (pu) at (-.3, -.3) {};
    \draw[sol] (aab) to [out=135, in=300] (pu);
    \draw[sol] (aac) to [out=0, in=205] (pu);
    \draw[pom] (aad) to [out=345, in=135] (pu);

    \node (pv) at (.3, -.3) {};
    \draw[sol] (abb) to [out=45, in=240] (pv);
    \draw[sol] (bbd) to [out=180, in=335] (pv);
    \draw[pom] (bbc) to [out=195, in=45] (pv);

    \node (pw) at (.3, .3) {};
    \draw[pom] (bcc) to [out=165, in=330] (pw);
    \draw[sol] (acc) to [out=180, in=25] (pw);
    \draw[pom] (ccd) to [out=300, in=135] (pw);

    \node (pz) at (-.3, .3) {};
    \draw[pom] (add) to [out=15, in=210] (pz);
    \draw[sol] (bdd) to [out=0, in=155] (pz);
    \draw[pom] (cdd) to [out=240, in=45] (pz);

    \node (q) at(0,0) {};
    \draw[pom] (pu) to [out=75, in=255] (q);
    \draw[pom] (pv) to [out=105, in=285] (q);
    \draw[sol] (pw) to [out=255, in=75] (q);
    \draw[sol] (pz) to [out=285, in=105] (q);
  \end{tikzpicture}
  \caption{No side or diagonal of the square was taken into $\mathcal{C}$}
\end{subfigure}
~
\begin{subfigure}{.32\textwidth}
  \begin{tikzpicture}
    \tikzset{every node/.style={draw,fill=black,circle,scale=.5,inner sep=0pt,outer sep=2pt}}
    \tikzstyle{pom}=[draw=black!60,thin]
    \node[vert]   (a) at (-1, -1) {};
    \node[vert]   (b) at (1, -1) {};
    \node[vert]   (c) at (1, 1) {};
    \node[vert]   (d) at (-1, 1) {};

    \node (aab) at (-.33,-1) {};
    \node (abb) at (.33, -1) {};

    \node (bbc) at (1, -.33) {};
    \node (bcc) at (1, .33) {};

    \node (ccd) at (.33, 1) {};
    \node (cdd) at (-.33, 1) {};

    \node (add) at (-1, .33) {};
    \node (aad) at (-1, -.33) {};

    \node (aac) at (-.7, -.7) {};
    \node (acc) at (.7, .7) {};
    \node (bbd) at (.7, -.7) {};
    \node (bdd) at (-.7, .7) {};

    \draw[pom] (a) -- (aab);
    \draw[pom] (abb) -- (b);
    \draw[pom] (b) -- (bbc);
    \draw[pom] (bcc) -- (c);
    \draw[sol] (c) -- (ccd);
    \draw[sol] (cdd) -- (d);
    \draw[pom] (d) -- (add);
    \draw[pom] (aad) -- (a);

    \draw[pom] (aab) -- (abb);
    \draw[sol] (bbc) -- (bcc);
    \draw[pom] (ccd) -- (cdd);
    \draw[sol] (add) -- (aad);

    \draw[pom] (a) -- (aac);
    \draw[pom] (b) -- (bbd);
    \draw[pom] (c) -- (acc);
    \draw[pom] (d) -- (bdd);

    \node (pu) at (-.3, -.3) {};
    \draw[sol] (aab) to [out=135, in=300] (pu);
    \draw[sol] (aac) to [out=0, in=205] (pu);
    \draw[pom] (aad) to [out=345, in=135] (pu);

    \node (pv) at (.3, -.3) {};
    \draw[sol] (abb) to [out=45, in=240] (pv);
    \draw[sol] (bbd) to [out=180, in=335] (pv);
    \draw[pom] (bbc) to [out=195, in=45] (pv);

    \node (pw) at (.3, .3) {};
    \draw[pom] (bcc) to [out=165, in=330] (pw);
    \draw[sol] (acc) to [out=180, in=25] (pw);
    \draw[pom] (ccd) to [out=300, in=135] (pw);

    \node (pz) at (-.3, .3) {};
    \draw[pom] (add) to [out=15, in=210] (pz);
    \draw[sol] (bdd) to [out=0, in=155] (pz);
    \draw[pom] (cdd) to [out=240, in=45] (pz);

    \node (q) at(0,0) {};
    \draw[pom] (pu) to [out=75, in=255] (q);
    \draw[pom] (pv) to [out=105, in=285] (q);
    \draw[sol] (pw) to [out=255, in=75] (q);
    \draw[sol] (pz) to [out=285, in=105] (q);
  \end{tikzpicture}
  \caption{$\mathcal{C}$ contains one side of the graph.}
\end{subfigure}
~
\begin{subfigure}{.32\textwidth}
  \begin{tikzpicture}
    \tikzset{every node/.style={draw,fill=black,circle,scale=.5,inner sep=0pt,outer sep=2pt}}
    \tikzstyle{pom}=[draw=black!60,thin]
    \node[vert]   (a) at (-1, -1) {};
    \node[vert]   (b) at (1, -1) {};
    \node[vert]   (c) at (1, 1) {};
    \node[vert]   (d) at (-1, 1) {};

    \node (aab) at (-.33,-1) {};
    \node (abb) at (.33, -1) {};

    \node (bbc) at (1, -.33) {};
    \node (bcc) at (1, .33) {};

    \node (ccd) at (.33, 1) {};
    \node (cdd) at (-.33, 1) {};

    \node (add) at (-1, .33) {};
    \node (aad) at (-1, -.33) {};

    \node (aac) at (-.7, -.7) {};
    \node (acc) at (.7, .7) {};
    \node (bbd) at (.7, -.7) {};
    \node (bdd) at (-.7, .7) {};

    \draw[pom] (a) -- (aab);
    \draw[pom] (abb) -- (b);
    \draw[sol] (b) -- (bbc);
    \draw[sol] (bcc) -- (c);
    \draw[pom] (c) -- (ccd);
    \draw[pom] (cdd) -- (d);
    \draw[sol] (d) -- (add);
    \draw[sol] (aad) -- (a);

    \draw[pom] (aab) -- (abb);
    \draw[pom] (bbc) -- (bcc);
    \draw[sol] (ccd) -- (cdd);
    \draw[pom] (add) -- (aad);

    \draw[pom] (a) -- (aac);
    \draw[pom] (b) -- (bbd);
    \draw[pom] (c) -- (acc);
    \draw[pom] (d) -- (bdd);

    \node (pu) at (-.3, -.3) {};
    \draw[sol] (aab) to [out=135, in=300] (pu);
    \draw[sol] (aac) to [out=0, in=205] (pu);
    \draw[pom] (aad) to [out=345, in=135] (pu);

    \node (pv) at (.3, -.3) {};
    \draw[sol] (abb) to [out=45, in=240] (pv);
    \draw[sol] (bbd) to [out=180, in=335] (pv);
    \draw[pom] (bbc) to [out=195, in=45] (pv);

    \node (pw) at (.3, .3) {};
    \draw[pom] (bcc) to [out=165, in=330] (pw);
    \draw[sol] (acc) to [out=180, in=25] (pw);
    \draw[pom] (ccd) to [out=300, in=135] (pw);

    \node (pz) at (-.3, .3) {};
    \draw[pom] (add) to [out=15, in=210] (pz);
    \draw[sol] (bdd) to [out=0, in=155] (pz);
    \draw[pom] (cdd) to [out=240, in=45] (pz);

    \node (q) at(0,0) {};
    \draw[pom] (pu) to [out=75, in=255] (q);
    \draw[pom] (pv) to [out=105, in=285] (q);
    \draw[sol] (pw) to [out=255, in=75] (q);
    \draw[sol] (pz) to [out=285, in=105] (q);
  \end{tikzpicture}
  \caption{$\mathcal{C}$ contains two opposite sides of $K$}
\end{subfigure}
~
\begin{subfigure}{.32\textwidth}
  \begin{tikzpicture}
    \tikzset{every node/.style={draw,fill=black,circle,scale=.5,inner sep=0pt,outer sep=2pt}}
    \tikzstyle{pom}=[draw=black!60,thin]
    \node[vert]   (a) at (-1, -1) {};
    \node[vert]   (b) at (1, -1) {};
    \node[vert]   (c) at (1, 1) {};
    \node[vert]   (d) at (-1, 1) {};

    \node (aab) at (-.33,-1) {};
    \node (abb) at (.33, -1) {};

    \node (bbc) at (1, -.33) {};
    \node (bcc) at (1, .33) {};

    \node (ccd) at (.33, 1) {};
    \node (cdd) at (-.33, 1) {};

    \node (add) at (-1, .33) {};
    \node (aad) at (-1, -.33) {};

    \node (aac) at (-.7, -.7) {};
    \node (acc) at (.7, .7) {};
    \node (bbd) at (.7, -.7) {};
    \node (bdd) at (-.7, .7) {};

    \draw[pom] (a) -- (aab);
    \draw[pom] (abb) -- (b);
    \draw[sol] (b) -- (bbc);
    \draw[sol] (bcc) -- (c);
    \draw[sol] (c) -- (ccd);
    \draw[sol] (cdd) -- (d);
    \draw[pom] (d) -- (add);
    \draw[pom] (aad) -- (a);

    \draw[pom] (aab) -- (abb);
    \draw[pom] (bbc) -- (bcc);
    \draw[pom] (ccd) -- (cdd);
    \draw[sol] (add) -- (aad);

    \draw[pom] (a) -- (aac);
    \draw[pom] (b) -- (bbd);
    \draw[pom] (c) -- (acc);
    \draw[pom] (d) -- (bdd);

    \node (pu) at (-.3, -.3) {};
    \draw[sol] (aab) to [out=135, in=300] (pu);
    \draw[sol] (aac) to [out=0, in=205] (pu);
    \draw[pom] (aad) to [out=345, in=135] (pu);

    \node (pv) at (.3, -.3) {};
    \draw[sol] (abb) to [out=45, in=240] (pv);
    \draw[sol] (bbd) to [out=180, in=335] (pv);
    \draw[pom] (bbc) to [out=195, in=45] (pv);

    \node (pw) at (.3, .3) {};
    \draw[pom] (bcc) to [out=165, in=330] (pw);
    \draw[sol] (acc) to [out=180, in=25] (pw);
    \draw[pom] (ccd) to [out=300, in=135] (pw);

    \node (pz) at (-.3, .3) {};
    \draw[pom] (add) to [out=15, in=210] (pz);
    \draw[sol] (bdd) to [out=0, in=155] (pz);
    \draw[pom] (cdd) to [out=240, in=45] (pz);

    \node (q) at(0,0) {};
    \draw[pom] (pu) to [out=75, in=255] (q);
    \draw[pom] (pv) to [out=105, in=285] (q);
    \draw[sol] (pw) to [out=255, in=75] (q);
    \draw[sol] (pz) to [out=285, in=105] (q);
  \end{tikzpicture}
  \caption{$\mathcal{C}$ contains two adjacent sides of the cycle $K$}
\end{subfigure}
~
\begin{subfigure}{.32\textwidth}
  \begin{tikzpicture}
    \tikzset{every node/.style={draw,fill=black,circle,scale=.5,inner sep=0pt,outer sep=2pt}}
    \tikzstyle{pom}=[draw=black!60,thin]
    \node[vert]   (a) at (-1, -1) {};
    \node[vert]   (b) at (1, -1) {};
    \node[vert]   (c) at (1, 1) {};
    \node[vert]   (d) at (-1, 1) {};

    \node (aab) at (-.33,-1) {};
    \node (abb) at (.33, -1) {};

    \node (bbc) at (1, -.33) {};
    \node (bcc) at (1, .33) {};

    \node (ccd) at (.33, 1) {};
    \node (cdd) at (-.33, 1) {};

    \node (add) at (-1, .33) {};
    \node (aad) at (-1, -.33) {};

    \node (aac) at (-.7, -.7) {};
    \node (acc) at (.7, .7) {};
    \node (bbd) at (.7, -.7) {};
    \node (bdd) at (-.7, .7) {};

    \draw[pom] (a) -- (aab);
    \draw[pom] (abb) -- (b);
    \draw[sol] (b) -- (bbc);
    \draw[sol] (bcc) -- (c);
    \draw[sol] (c) -- (ccd);
    \draw[sol] (cdd) -- (d);
    \draw[sol] (d) -- (add);
    \draw[sol] (aad) -- (a);

    \draw[pom] (aab) -- (abb);
    \draw[pom] (bbc) -- (bcc);
    \draw[pom] (ccd) -- (cdd);
    \draw[pom] (add) -- (aad);

    \draw[pom] (a) -- (aac);
    \draw[pom] (b) -- (bbd);
    \draw[pom] (c) -- (acc);
    \draw[pom] (d) -- (bdd);

    \node (pu) at (-.3, -.3) {};
    \draw[sol] (aab) to [out=135, in=300] (pu);
    \draw[sol] (aac) to [out=0, in=205] (pu);
    \draw[pom] (aad) to [out=345, in=135] (pu);

    \node (pv) at (.3, -.3) {};
    \draw[sol] (abb) to [out=45, in=240] (pv);
    \draw[sol] (bbd) to [out=180, in=335] (pv);
    \draw[pom] (bbc) to [out=195, in=45] (pv);

    \node (pw) at (.3, .3) {};
    \draw[pom] (bcc) to [out=165, in=330] (pw);
    \draw[sol] (acc) to [out=180, in=25] (pw);
    \draw[pom] (ccd) to [out=300, in=135] (pw);

    \node (pz) at (-.3, .3) {};
    \draw[pom] (add) to [out=15, in=210] (pz);
    \draw[sol] (bdd) to [out=0, in=155] (pz);
    \draw[pom] (cdd) to [out=240, in=45] (pz);

    \node (q) at(0,0) {};
    \draw[pom] (pu) to [out=75, in=255] (q);
    \draw[pom] (pv) to [out=105, in=285] (q);
    \draw[sol] (pw) to [out=255, in=75] (q);
    \draw[sol] (pz) to [out=285, in=105] (q);
  \end{tikzpicture}
  \caption{Three sides of the square are taken into $\mathcal{C}$.}
\end{subfigure}
~
\begin{subfigure}{.32\textwidth}
  \begin{tikzpicture}
    \tikzset{every node/.style={draw,fill=black,circle,scale=.5,inner sep=0pt,outer sep=2pt}}
    \tikzstyle{pom}=[draw=black!60,thin]
    \node[vert]   (a) at (-1, -1) {};
    \node[vert]   (b) at (1, -1) {};
    \node[vert]   (c) at (1, 1) {};
    \node[vert]   (d) at (-1, 1) {};

    \node (aab) at (-.33,-1) {};
    \node (abb) at (.33, -1) {};

    \node (bbc) at (1, -.33) {};
    \node (bcc) at (1, .33) {};

    \node (ccd) at (.33, 1) {};
    \node (cdd) at (-.33, 1) {};

    \node (add) at (-1, .33) {};
    \node (aad) at (-1, -.33) {};

    \node (aac) at (-.7, -.7) {};
    \node (acc) at (.7, .7) {};
    \node (bbd) at (.7, -.7) {};
    \node (bdd) at (-.7, .7) {};

    \draw[pom] (a) -- (aab);
    \draw[pom] (abb) -- (b);
    \draw[pom] (b) -- (bbc);
    \draw[pom] (bcc) -- (c);
    \draw[pom] (c) -- (ccd);
    \draw[pom] (cdd) -- (d);
    \draw[pom] (d) -- (add);
    \draw[pom] (aad) -- (a);

    \draw[pom] (aab) -- (abb);
    \draw[sol] (bbc) -- (bcc);
    \draw[pom] (ccd) -- (cdd);
    \draw[sol] (add) -- (aad);

    \draw[sol] (a) -- (aac);
    \draw[pom] (b) -- (bbd);
    \draw[sol] (c) -- (acc);
    \draw[pom] (d) -- (bdd);

    \node (pu) at (-.3, -.3) {};
    \draw[sol] (aab) to [out=135, in=300] (pu);
    \draw[pom] (aac) to [out=0, in=205] (pu);
    \draw[pom] (aad) to [out=345, in=135] (pu);

    \node (pv) at (.3, -.3) {};
    \draw[sol] (abb) to [out=45, in=240] (pv);
    \draw[sol] (bbd) to [out=180, in=335] (pv);
    \draw[pom] (bbc) to [out=195, in=45] (pv);

    \node (pw) at (.3, .3) {};
    \draw[pom] (bcc) to [out=165, in=330] (pw);
    \draw[pom] (acc) to [out=180, in=25] (pw);
    \draw[sol] (ccd) to [out=300, in=135] (pw);

    \node (pz) at (-.3, .3) {};
    \draw[pom] (add) to [out=15, in=210] (pz);
    \draw[sol] (bdd) to [out=0, in=155] (pz);
    \draw[sol] (cdd) to [out=240, in=45] (pz);

    \node (q) at(0,0) {};
    \draw[sol] (pu) to [out=75, in=255] (q);
    \draw[pom] (pv) to [out=105, in=285] (q);
    \draw[sol] (pw) to [out=255, in=75] (q);
    \draw[pom] (pz) to [out=285, in=105] (q);
  \end{tikzpicture}
  \caption{$\mathcal{C}$ contains one diagonal of $K$.}
\end{subfigure}
~
\begin{subfigure}{.32\textwidth}
  \begin{tikzpicture}
    \tikzset{every node/.style={draw,fill=black,circle,scale=.5,inner sep=0pt,outer sep=2pt}}
    \tikzstyle{pom}=[draw=black!60,thin]
    \node[vert]   (a) at (-1, -1) {};
    \node[vert]   (b) at (1, -1) {};
    \node[vert]   (c) at (1, 1) {};
    \node[vert]   (d) at (-1, 1) {};

    \node (aab) at (-.33,-1) {};
    \node (abb) at (.33, -1) {};

    \node (bbc) at (1, -.33) {};
    \node (bcc) at (1, .33) {};

    \node (ccd) at (.33, 1) {};
    \node (cdd) at (-.33, 1) {};

    \node (add) at (-1, .33) {};
    \node (aad) at (-1, -.33) {};

    \node (aac) at (-.7, -.7) {};
    \node (acc) at (.7, .7) {};
    \node (bbd) at (.7, -.7) {};
    \node (bdd) at (-.7, .7) {};

    \draw[pom] (a) -- (aab);
    \draw[pom] (abb) -- (b);
    \draw[pom] (b) -- (bbc);
    \draw[pom] (bcc) -- (c);
    \draw[pom] (c) -- (ccd);
    \draw[pom] (cdd) -- (d);
    \draw[sol] (d) -- (add);
    \draw[sol] (aad) -- (a);

    \draw[pom] (aab) -- (abb);
    \draw[sol] (bbc) -- (bcc);
    \draw[pom] (ccd) -- (cdd);
    \draw[pom] (add) -- (aad);

    \draw[sol] (a) -- (aac);
    \draw[pom] (b) -- (bbd);
    \draw[sol] (c) -- (acc);
    \draw[pom] (d) -- (bdd);

    \node (pu) at (-.3, -.3) {};
    \draw[sol] (aab) to [out=135, in=300] (pu);
    \draw[pom] (aac) to [out=0, in=205] (pu);
    \draw[pom] (aad) to [out=345, in=135] (pu);

    \node (pv) at (.3, -.3) {};
    \draw[sol] (abb) to [out=45, in=240] (pv);
    \draw[sol] (bbd) to [out=180, in=335] (pv);
    \draw[pom] (bbc) to [out=195, in=45] (pv);

    \node (pw) at (.3, .3) {};
    \draw[pom] (bcc) to [out=165, in=330] (pw);
    \draw[pom] (acc) to [out=180, in=25] (pw);
    \draw[sol] (ccd) to [out=300, in=135] (pw);

    \node (pz) at (-.3, .3) {};
    \draw[pom] (add) to [out=15, in=210] (pz);
    \draw[sol] (bdd) to [out=0, in=155] (pz);
    \draw[sol] (cdd) to [out=240, in=45] (pz);

    \node (q) at(0,0) {};
    \draw[sol] (pu) to [out=75, in=255] (q);
    \draw[pom] (pv) to [out=105, in=285] (q);
    \draw[sol] (pw) to [out=255, in=75] (q);
    \draw[pom] (pz) to [out=285, in=105] (q);
  \end{tikzpicture}
  \caption{A diagonal and a side edge of $K$ are in $\mathcal{C}$.}
\end{subfigure}
~
\begin{subfigure}{.32\textwidth}
  \begin{tikzpicture}
    \tikzset{every node/.style={draw,fill=black,circle,scale=.5,inner sep=0pt,outer sep=2pt}}
    \tikzstyle{pom}=[draw=black!60,thin]
    \node[vert]   (a) at (-1, -1) {};
    \node[vert]   (b) at (1, -1) {};
    \node[vert]   (c) at (1, 1) {};
    \node[vert]   (d) at (-1, 1) {};

    \node (aab) at (-.33,-1) {};
    \node (abb) at (.33, -1) {};

    \node (bbc) at (1, -.33) {};
    \node (bcc) at (1, .33) {};

    \node (ccd) at (.33, 1) {};
    \node (cdd) at (-.33, 1) {};

    \node (add) at (-1, .33) {};
    \node (aad) at (-1, -.33) {};

    \node (aac) at (-.7, -.7) {};
    \node (acc) at (.7, .7) {};
    \node (bbd) at (.7, -.7) {};
    \node (bdd) at (-.7, .7) {};

    \draw[pom] (a) -- (aab);
    \draw[pom] (abb) -- (b);
    \draw[sol] (b) -- (bbc);
    \draw[sol] (bcc) -- (c);
    \draw[pom] (c) -- (ccd);
    \draw[pom] (cdd) -- (d);
    \draw[sol] (d) -- (add);
    \draw[sol] (aad) -- (a);

    \draw[pom] (aab) -- (abb);
    \draw[pom] (bbc) -- (bcc);
    \draw[pom] (ccd) -- (cdd);
    \draw[pom] (add) -- (aad);

    \draw[sol] (a) -- (aac);
    \draw[pom] (b) -- (bbd);
    \draw[sol] (c) -- (acc);
    \draw[pom] (d) -- (bdd);

    \node (pu) at (-.3, -.3) {};
    \draw[sol] (aab) to [out=135, in=300] (pu);
    \draw[pom] (aac) to [out=0, in=205] (pu);
    \draw[pom] (aad) to [out=345, in=135] (pu);

    \node (pv) at (.3, -.3) {};
    \draw[sol] (abb) to [out=45, in=240] (pv);
    \draw[sol] (bbd) to [out=180, in=335] (pv);
    \draw[pom] (bbc) to [out=195, in=45] (pv);

    \node (pw) at (.3, .3) {};
    \draw[pom] (bcc) to [out=165, in=330] (pw);
    \draw[pom] (acc) to [out=180, in=25] (pw);
    \draw[sol] (ccd) to [out=300, in=135] (pw);

    \node (pz) at (-.3, .3) {};
    \draw[pom] (add) to [out=15, in=210] (pz);
    \draw[sol] (bdd) to [out=0, in=155] (pz);
    \draw[sol] (cdd) to [out=240, in=45] (pz);

    \node (q) at(0,0) {};
    \draw[sol] (pu) to [out=75, in=255] (q);
    \draw[pom] (pv) to [out=105, in=285] (q);
    \draw[sol] (pw) to [out=255, in=75] (q);
    \draw[pom] (pz) to [out=285, in=105] (q);
  \end{tikzpicture}
  \caption{$\mathcal{C}$ contains one diagonal and two opposite side edges
    of~$K$.}
\end{subfigure}
~
\begin{subfigure}{.32\textwidth}
  \begin{tikzpicture}
    \tikzset{every node/.style={draw,fill=black,circle,scale=.5,inner sep=0pt,outer sep=2pt}}
    \tikzstyle{pom}=[draw=black!60,thin]
    \node[vert]   (a) at (-1, -1) {};
    \node[vert]   (b) at (1, -1) {};
    \node[vert]   (c) at (1, 1) {};
    \node[vert]   (d) at (-1, 1) {};

    \node (aab) at (-.33,-1) {};
    \node (abb) at (.33, -1) {};

    \node (bbc) at (1, -.33) {};
    \node (bcc) at (1, .33) {};

    \node (ccd) at (.33, 1) {};
    \node (cdd) at (-.33, 1) {};

    \node (add) at (-1, .33) {};
    \node (aad) at (-1, -.33) {};

    \node (aac) at (-.7, -.7) {};
    \node (acc) at (.7, .7) {};
    \node (bbd) at (.7, -.7) {};
    \node (bdd) at (-.7, .7) {};

    \draw[pom] (a) -- (aab);
    \draw[pom] (abb) -- (b);
    \draw[pom] (b) -- (bbc);
    \draw[pom] (bcc) -- (c);
    \draw[pom] (c) -- (ccd);
    \draw[pom] (cdd) -- (d);
    \draw[pom] (d) -- (add);
    \draw[pom] (aad) -- (a);

    \draw[pom] (aab) -- (abb);
    \draw[pom] (bbc) -- (bcc);
    \draw[sol] (ccd) -- (cdd);
    \draw[pom] (add) -- (aad);

    \draw[sol] (a) -- (aac);
    \draw[sol] (b) -- (bbd);
    \draw[sol] (c) -- (acc);
    \draw[sol] (d) -- (bdd);

    \node (pu) at (-.3, -.3) {};
    \draw[sol] (aab) to [out=135, in=300] (pu);
    \draw[pom] (aac) to [out=0, in=205] (pu);
    \draw[sol] (aad) to [out=345, in=135] (pu);

    \node (pv) at (.3, -.3) {};
    \draw[sol] (abb) to [out=45, in=240] (pv);
    \draw[pom] (bbd) to [out=180, in=335] (pv);
    \draw[sol] (bbc) to [out=195, in=45] (pv);

    \node (pw) at (.3, .3) {};
    \draw[sol] (bcc) to [out=165, in=330] (pw);
    \draw[pom] (acc) to [out=180, in=25] (pw);
    \draw[pom] (ccd) to [out=300, in=135] (pw);

    \node (pz) at (-.3, .3) {};
    \draw[sol] (add) to [out=15, in=210] (pz);
    \draw[pom] (bdd) to [out=0, in=155] (pz);
    \draw[pom] (cdd) to [out=240, in=45] (pz);

    \node (q) at(0,0) {};
    \draw[pom] (pu) to [out=75, in=255] (q);
    \draw[pom] (pv) to [out=105, in=285] (q);
    \draw[sol] (pw) to [out=255, in=75] (q);
    \draw[sol] (pz) to [out=285, in=105] (q);
  \end{tikzpicture}
  \caption{Two diagonals of $K$ are taken into $\mathcal{C}$.}
\end{subfigure}
~
\begin{subfigure}{.32\textwidth}
  \begin{tikzpicture}
    \tikzset{every node/.style={draw,fill=black,circle,scale=.5,inner sep=0pt,outer sep=2pt}}
    \tikzstyle{pom}=[draw=black!60,thin]
    \node[vert]   (a) at (-1, -1) {};
    \node[vert]   (b) at (1, -1) {};
    \node[vert]   (c) at (1, 1) {};
    \node[vert]   (d) at (-1, 1) {};

    \node (aab) at (-.33,-1) {};
    \node (abb) at (.33, -1) {};

    \node (bbc) at (1, -.33) {};
    \node (bcc) at (1, .33) {};

    \node (ccd) at (.33, 1) {};
    \node (cdd) at (-.33, 1) {};

    \node (add) at (-1, .33) {};
    \node (aad) at (-1, -.33) {};

    \node (aac) at (-.7, -.7) {};
    \node (acc) at (.7, .7) {};
    \node (bbd) at (.7, -.7) {};
    \node (bdd) at (-.7, .7) {};

    \draw[pom] (a) -- (aab);
    \draw[pom] (abb) -- (b);
    \draw[pom] (b) -- (bbc);
    \draw[pom] (bcc) -- (c);
    \draw[sol] (c) -- (ccd);
    \draw[sol] (cdd) -- (d);
    \draw[pom] (d) -- (add);
    \draw[pom] (aad) -- (a);

    \draw[pom] (aab) -- (abb);
    \draw[pom] (bbc) -- (bcc);
    \draw[pom] (ccd) -- (cdd);
    \draw[pom] (add) -- (aad);

    \draw[sol] (a) -- (aac);
    \draw[sol] (b) -- (bbd);
    \draw[sol] (c) -- (acc);
    \draw[sol] (d) -- (bdd);

    \node (pu) at (-.3, -.3) {};
    \draw[sol] (aab) to [out=135, in=300] (pu);
    \draw[pom] (aac) to [out=0, in=205] (pu);
    \draw[sol] (aad) to [out=345, in=135] (pu);

    \node (pv) at (.3, -.3) {};
    \draw[sol] (abb) to [out=45, in=240] (pv);
    \draw[pom] (bbd) to [out=180, in=335] (pv);
    \draw[sol] (bbc) to [out=195, in=45] (pv);

    \node (pw) at (.3, .3) {};
    \draw[sol] (bcc) to [out=165, in=330] (pw);
    \draw[pom] (acc) to [out=180, in=25] (pw);
    \draw[pom] (ccd) to [out=300, in=135] (pw);

    \node (pz) at (-.3, .3) {};
    \draw[sol] (add) to [out=15, in=210] (pz);
    \draw[pom] (bdd) to [out=0, in=155] (pz);
    \draw[pom] (cdd) to [out=240, in=45] (pz);

    \node (q) at(0,0) {};
    \draw[pom] (pu) to [out=75, in=255] (q);
    \draw[pom] (pv) to [out=105, in=285] (q);
    \draw[sol] (pw) to [out=255, in=75] (q);
    \draw[sol] (pz) to [out=285, in=105] (q);
  \end{tikzpicture}
  \caption{Two diagonals and a side of $K$ are in $\mathcal{C}$.}
\end{subfigure}
\end{center}

    \caption{Proof of the Lemma \ref{lem:maxtsp_g2compute_nobreak_sqkite}. For
    every selection of the edges and diagonals of $K$ in the cycle cover
    $\mathcal{C}$ we are showing, how to select edges of the gadget
    $\mathcal{G}_K$, to realize the  cycle cover in the gadgets-modified graph.}
    \label{fig:maxtsp_g2compute_nobreak_sqkite}
  \end{figure}
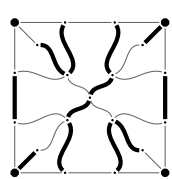
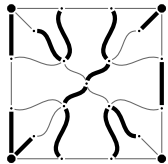
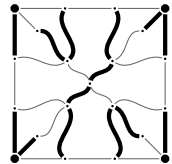
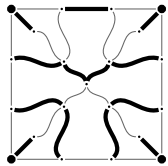

  In turn, for the 4-kite it will turn out, that our gadgets not only block
  selecting a length-4 cycle into the b-matching, but also prevent it from
  containing a length-3 cycle built on three vertices of the 4-kite.
  \begin{lemma}
    \label{lem:maxtsp_g2compute_nobreak_sqkite}
    Let $K \in \mathcal{C}_1$ be a 4-kite in $G_1$. Let $\mathcal{C}$ be a cycle cover of $G$ not containing any length-4 or length-3 cycle built on the vertices of $K$ as one of its cycles. There exists a selection of edges, that is compliant with the cycle cover $\mathcal{C}$, such that every vertex $v \in \mathcal{G}_K$ has exactly $b(v)$ adjacent edges in the selection (so the selection forms a b-matching). The weight of the b-matching is equal to that of $\mathcal{C}$.
  \end{lemma}
  \dowod
    Similarly to the proof of Lemma \ref{lem:maxtsp_g2compute_trikite}, we
    need to look into all the possible interactions of the cycle cover
    $\mathcal{C}$ with the edges of $K$ (together with its diagonals). For every
    such option, we will show, how to expand it into a compliant b-matching. The
    analysis is presented in the
    Figure \ref{fig:maxtsp_g2compute_nobreak_sqkite}
  \koniec

\end{document}